\documentclass[alpha-refs]{wiley-article}
\usepackage{siunitx}
\usepackage[utf8]{inputenc}
\usepackage{amsmath}
\usepackage{amssymb}
\usepackage{bm}
\usepackage{cases}
\usepackage{graphicx}
\usepackage[super]{nth}
\usepackage{mathtools}
\usepackage{xcolor}
\usepackage{subcaption}
\usepackage{wrapfig}
\usepackage[hyphens]{url}
\usepackage{hyperref}
\papertype{Original Article}
\paperfield{Journal Section}

\newcommand{\dd}{\mathrm{d}}
\newcommand{\ep}{\mathrm{e}}
\newcommand{\Rey}{\mathrm{Re}}
\newcommand{\Ri}{\mathrm{Ri}}

\title{On the Holton-Lindzen-Plumb model for mean flow reversals in stratified fluids}

\author[1\authfn{1}]{Antoine Renaud}
\author[2\authfn{1}]{Antoine Venaille}

\affil[1]{School  of  Mathematics  and  Maxwell  Institute  for  Mathematical  Sciences,  University  of  Edinburgh, King’s Buildings, Edinburgh EH9 3FD, United Kingdom}
\affil[2]{Univ Lyon, Ens de Lyon, Univ Claude Bernard, CNRS, Laboratoire de Physique, F-69342 Lyon, France}

\corremail{antoine.renaud@ed.ac.uk}

\runningauthor{A. Renaud \& A. Venaille}

\begin{document}

\maketitle
\begin{abstract}
The Holton-Lindzen-Plumb model describes the spontaneous emergence of mean flow reversals in stratified fluids. It has played a central role in understanding the quasi-biennial oscillation of equatorial winds in Earth's stratosphere and has arguably become a linchpin of wave-mean flow interaction theory in geophysical and astrophysical fluid dynamics. The derivation of the model's equation from primitive equations follows from several assumptions, including quasi-linear approximations, WKB expansion of the wavefield, simplifications of boundary layer terms, among others. Starting from the two-dimensional, non-rotating, Boussinesq equations, we present in this paper a self-consistent derivation of the Holton-Lindzen-Plumb model and show the existence of a distinguished limit for which all approximations remains valid. We furthermore discuss the important role of boundary conditions, and the relevance of this model to describe secondary bifurcations associated with a quasi-periodic route to chaos.
\keywords{wave-mean flow interaction, stratified fluid, quasi-biennial oscillation, internal waves, streaming}
\end{abstract}

\section{Introduction\label{sec:Intro}}

 Roughly every 14 months, winds in the equatorial stratosphere reverse direction, alternating between westward and eastward phases. This phenomenon called quasi-biennial oscillations (QB0) is arguably the clearest example of spontaneously generated low-frequency periodic phenomenon in geophysical flows, i.e. without direct link with astronomical forcing such as the seasonal cycle \citep{Baldwin2001}. \cite{Lindzen1968} elucidated at the end of the sixties the basic mechanisms underlying this phenomenon and proposed in the early seventies a simplified model for the reversals \citep{Holton1972}. They explained the interplay between waves, dissipative effects and mean flows in the equatorial stratosphere, with an emphasis on the role of planetary Yanai and Kelvin waves \citep{lindzen1971equatorial}. Building on their model,  \cite{Plumb1977}  isolated a few years later the minimal ingredients required to observe the spontaneous generation of mean flow reversals in stratified fluid: a horizontally periodic domain filled with a stratified fluid forced at the bottom by a source of waves with a horizontal phase speed of opposite sign. The sufficiency of these basic elements has been successfully demonstrated with a now celebrated laboratory experiment \citep{Plumb1978}. The spontaneous generation of low-frequency oscillations in the Holton-Lindzen-Plumb model describes the spontaneous emergence of mean flow reversals in stratified fluids. It has played a central role in understanding the quasi-biennial oscillation of equatorial winds in Earth's stratosphere and has arguably become a linchpin of wave-mean flow interaction theory in geophysical and astrophysical fluid dynamics. The derivation of the model's equation from primitive equations follows from several assumptions, including quasi-linear approximations, WKB expansion of the wavefield, simplifications of boundary layer terms, among others. Starting from the two-dimensional, non-rotating, Boussinesq equations, we present in this paper a self-consistent derivation of the Holton-Lindzen-Plumb model and show the existence of a distinguished limit for which all approximations remains valid. We furthermore discuss the important role of boundary conditions, and the relevance of this model to describe secondary bifurcations associated with a quasi-periodic route to chaos.e experiments was interpreted with a partial integrodifferential equation for the velocity that is now presented in most geophysical fluid dynamics textbooks, and that we call the Holton-Lindzen-Plumb's equation or model. In nondimensional settings, the equation reads
 \begin{equation}\label{eq:Holton-Lindzen-Plumb_intro}
\partial_{T}U- \Rey ^{-1}\partial_Z^2U=-\partial_Z\left(\exp\left\{-\int_0^Z\frac{\mathrm{d}Z'}{\left(1-U\right)^2}\right\}-\exp\left\{-\int_0^Z\frac{\mathrm{d}Z'}{\left(1+U\right)^2}\right\}\right),
\end{equation} 
with appropriate boundary conditions. The model predicts the evolution of the averaged zonal velocity fields $U(Z,t)$ controlled solely by the Reynolds number $\mathrm{Re}$. A historical perspective on the development of such low-dimensional QBO models is provided by \cite{lindzen1987development}. We revisit in this paper the model's derivation, paying particular attention to the underlying hypothesis, most of them (but not all) being listed in the original paper \citep{Plumb1975,Plumb1977}; we discuss the relevant nondimensional parameters of the problem and show the existence of a distinguished limit for which the model is self-consistent. This analysis highlights the important role of boundary conditions.

 The motivation for this work comes from a revival of interest in QBO-like phenomena over the last few years. First, an unexpected periodicity disruption of QBO on Earth was reported in 2016 \citep{newman2016anomalous,osprey2016unexpected}, triggering debates on the origin of this effect \citep{dunkerton2016quasi}. Second, QBO like phenomena have been reported in other planetary atmosphere \citep{Dowling2008,Read2018}, and is suspected to occur in stably stratified layers in stars \citep{Mcintyre1994,kim2001gravity,rogers2006angular,rogers2008non,Showman2018}.  Third, fluid dynamicists have shed new light on the interplay between waves, mean flow and sometimes turbulence in stratified fluids: the nature of the bifurcation towards an oscillating state in Holton-Lindzen-Plumb's model when the control parameter $\mathrm{Re}$ is varied has been elucidated and tested against laboratory experiments \citep{yoden1988new,semin2018nonlinear}; the possibility for synchronization or phase-locking with a seasonal cycle have been investigated within Holton-Lindzen-Plumb's framework \citep{rajendran2015}; emergence of low-frequency mean-flow reversals in a stably stratified layer forced by a turbulent layer have been reported in direct numerical simulations \citep{Couston2018}; and secondary bifurcations with a quasi-periodic route to chaos in mean flow reversals have been reported both in Holton-Lindzen-Plumb's model \citep{kim2001gravity,renaud2019periodicity} and direct numerical simulations \citep{renaud2019periodicity}. This large number of studies involving Holton-Lindzen-Plumb's equation in different contexts has been a strong incentive to take a closer look at the derivation of this model and its limitations.

The paper is organised as follows.  In the second section, we list the series of assumptions that leads to Holton-Lindzen-Plumb's model, starting from the non-rotating Boussinesq equations in a simplified two-dimensional geometry, paying particular attention to the nondimensional parameters of the problem and to the physical mechanisms that govern wave-mean flow interactions in this context. The third section recalls the phenomenology of the mean flow evolution in the Holton-Lindzen-Plumb's model, considering simple limiting cases, namely forcing by either a single propagating wave at the bottom or by two-counter propagating waves. In particular, we propose an analytic form for stationary states in the case where the flow is forced by a single monochromatic wave at the bottom. This analytic form happens to be also a useful guide to interpret the mean flow profiles in the case of forcing with two counter-propagating waves. These observations and the scaling derived from the analytic profile are used in Appendix A to justify \textit{a posteriori} the derivation of Holton-Lindzen-Plumb's model and to show the existence of a distinguished limit for which the derivation is self-consistent. This is the main result of this paper. In the fourth section, we review the current understanding of the QBO bifurcations, the role of symmetries, and the possibilities of additional bifurcations. We emphasise in this section the central role of bottom boundary conditions, that had up to now largely been ignored: changing this condition from no slip to free slip invalidates self-consistency of the model derivation and favours the transition to chaos.

\section{From Boussinesq equations to Holton-Lindzen-Plumb model\label{sec:Holton-Lindzen-Plumbmodel}}
In this section, we propose a derivation of the Holton-Lindzen-Plumb's model, starting from the non-rotating Boussinesq equations. The first subsection introduces the primary model's equations and geometry. Starting from this minimal bedrock, the second subsection describes the quasilinear approximation, and the third subsection presents the approximations leading to a closure for Reynolds stresses. The approximations are carefully listed and the self-consistency of the model is checked a posteriori in appendix \ref{sec:SelfConsHolton-Lindzen-Plumb}. En route, this section also highlights the essential physical mechanisms at play.
    \subsection{Primitive set of equations \label{ssec:BaseModel}}
    We consider a two-dimensional vertical slice of fluid, periodic in the zonal direction with period $L$, and semi-infinite in the upward direction. The zonal and vertical coordinates are labelled by $x$ and $z$ and are associated with the unit vectors $\bm{e}_{x}$ and $\bm{e}_{z}$ respectively. We consider a linear stratification profile with buoyancy frequency $N$. The energy is dissipated by two processes: i) the viscous damping with kinematic viscosity $\nu$ ii) the linear damping of buoyancy disturbances with rate $\gamma$. The fluid motion is governed by the 2D Navier\--Stokes equations under the Boussinesq approximation
    \begin{subnumcases}{\label{eq:BoussModel}}
        \partial_{t}\bm{u}+\left(\bm{u}\cdot \nabla\right) \bm{u}&$=-\nabla\Phi+b\bm{e}_z+\nu\nabla^{2} \bm{u}$,\label{eq:BoussModel_1}\\
        \partial_{t}b+\bm{u}\cdot\nabla b+N^{2}w&$=-\gamma b$,\label{eq:BoussModel_2}\\
        \nabla\cdot\bm{u}&$=0$\label{eq:BoussModel_3},
    \end{subnumcases}
    where $\bm{u}=u\bm{e}_{x}+w\bm{e}_{z}$ is the two-dimensional velocity field, $\nabla=(\partial_{x}, \partial_{z})$ is the gradient operator, $\Phi$ is the pressure potential and $b$ is the buoyancy anomaly. Respectively, equations \eqref{eq:BoussModel_1}, \eqref{eq:BoussModel_2} and \eqref{eq:BoussModel_3} correspond to the momentum, buoyancy and mass conservation.

    The fluid is forced by the vertical undulations of a bottom boundary \-- periodic in time with period $T$. We denote $h\left(x,t\right)$ the deviation of the boundary from its mean position $z=0$. The boundary conditions are discussed below.
    
    \subsubsection*{Wave-mean flow decomposition\label{sssec:WMF}}
    We split the dynamics into a mean and wave part by averaging over the spatial $x$-wise periodicity of the domain and the temporal periodicity of the forcing
    \begin{equation}
        u\left(x,z,t\right)=\overline{u}\left(z,t\right)+u'\left(x,z,t\right)\quad\text{with}\quad \overline{u}\left(z,t\right)=\frac{1}{L}\int_{0}^{L}\frac{1}{T}\int_{0}^{T}\mathrm{d}x\mathrm{d}\tau\,u\left(x,z,t+\tau\right).\label{eq:WMF_decomp}
    \end{equation} 
    Here, $\overline{u}$ is the mean part while $u'$ is the wave part. Our choice to average over time as well as space will be useful to filter out high frequency mean flow oscillations. This average commutes with any derivative. Consequently, averaging Eq. \eqref{eq:BoussModel_3} constrains the mean vertical velocity to be $z$\--independent: $\partial_z{\overline{w}}=0$. Assuming no mass flux from below then leads to $\overline{w}=0$.
    
    \subsubsection*{Mean-flow equation}
    Averaging the horizontal projection of the momentum equation \eqref{eq:BoussModel_1} leads to the \textit{mean-flow equation}
    \begin{equation}
        \partial_{t}\overline{u}-\nu\partial_{z}^{2}\overline{u}=-\partial_{z}\overline{u'w'}. \label{eq:MeanFlowEq}
    \end{equation}
   The mean flow $\overline{u}$ is forced the vertical divergence of the Reynolds stress component or mean upward momentum flux $\overline{u'w'}$. Wave attenuation generates a mean flow through this forcing term. This phenomenon is often referred to as \emph{streaming} in fluid mechanics. The classical book of Lighthill presents this subject with an emphasis on analogies between acoustics and internal waves \citep{Lighthill1978}. A general introduction to wave mean-flow interactions theories is given in \citep{staquet2005internal,Buhler2009}. In the following, we will sometimes refer to the momentum flux divergence as the \emph{streaming force}.
   
   An exact computation of the momentum flux $\overline{u'w'}$ would require solving the primitive equations \eqref{eq:BoussModel} and do the averaging afterwards to obtain the mean flow. This task, intractable analytically, requires costly direct numerical simulations. Another approach consists in parameterizing the momentum flux $\overline{u'w'}$ to close the mean flow equation \eqref{eq:MeanFlowEq}. In their pioneering work Holton, Lindzen and Plumb built the first parameterizations reflecting the minimal ingredients required to obtain mean-flow reversals and jointly brought to light today's prevalent physical understanding of the quasi-biennial oscillation of equatorial winds in the Earth stratosphere.
   
    \subsubsection*{Wave equations}
    Subtracting the averaged equations from the dynamical equations \eqref{eq:BoussModel} yields the nonlinear \textit{wave equations}
    \begin{subnumcases}{\label{eq:WaveEq}}
        \partial_{t}\bm{u}'+\overline{u}\partial_{x}\bm{u}'+w'\partial_{z}\overline{u}\bm{e}_{x}&$=-\nabla\Phi'+b'\bm{e}_z+\nu\nabla^{2} \bm{u}'-\left(\left(\bm{u}'\cdot\nabla\right)\bm{u}^ {\prime}\right)'$,\label{eq:WaveEq_1}\\
        \partial_{t}b'+\overline{u}\partial_{x}b'+\left(N^{2}+\partial_{z}\overline{b}\right)w'&$=-\gamma b'-\left(\bm{u}'\cdot\nabla b'\right)'$,\label{eq:WaveEq_2}\\
        \nabla\cdot\bm{u}'&$=0$\label{eq:WaveEq_3}.
    \end{subnumcases}
    
    \subsubsection*{Boundary conditions\label{sssec:BCCond}}
    We consider a no-slip condition at the bottom boundary\footnote{ The horizontal average in \eqref{eq:WMF_decomp} is ill-defined close to the curvy bottom boundary. This issue is bypassed by Taylor expanding the bottom boundary condition \eqref{eq:BC_cond} with respect to $h$ such that the fields are defined up to the flat boundary $z=0$.} whose motion is assumed to be purely vertical
    \begin{equation}
     \bm{u}|_{z=h}=\partial_{t}h_b\,\bm{e}_{z}.\label{eq:BC_cond}
    \end{equation}
    and a free-slip boundary condition at infinity
     \begin{equation}
     \partial_z u|_{z=+\infty}=0\quad,\quad\partial_x w|_{z=+\infty}=0.\label{eq:BC_cond_inf}
    \end{equation}
    Such bottom boundary condition suites well to laboratory experiment contexts where internal waves are generated by solid membrane oscillations \citep{Plumb1978,Otobe1998,semin2018nonlinear}. In the stratospheric context, the oscillating bottom boundary mimics the tropopause height variations forced by the deep convection in the equatorial troposphere. In the atmospheric case, it is not obvious that the no-slip condition at the bottom is the relevant choice. It is nevertheless the most commonly used boundary condition in the literature (see e.g. \cite{Plumb1977,rajendran2015} among others). We discuss the case of free-slip boundary condition and its implications in appendix \ref{sec:free-slip}.

\subsubsection*{Characteristic scales and nondimensional parameters \label{ssec:NonDimParam}}

The parameters of the problem are the buoyancy frequency $N$, the buoyancy damping rate $\gamma$, the viscosity $\nu$ and the characteristic amplitude  $h$, angular frequency $\omega$ (always considered positive) and horizontal wavenumber $k$ of the bottom undulation.  These $6$ parameters involve only time and space units. This corresponds to $4$ nondimensional independent parameters\footnote{We assume that the domain is semi-infinite in the vertical direction thus preventing non-trivial effects related to wave-reflections, such as the emergence of internal wave attractors. We also neglect buoyancy diffusivity, which amounts to assume an infinite Prandtl number.}:
\begin{equation}\label{eq:NondimensionalNumbers}
    hk\quad,\quad\frac{\omega}{N}\quad,\quad\frac{\gamma}{\omega}\quad\text{and}\quad \Rey\equiv \frac{\omega^2 h^2}{2\gamma \nu}
\end{equation}
where $\Rey$ is a Reynolds number that will appear naturally in the model derivation. In the following, we consider a distinguished limit where $hk$, $\omega/N$ and $\gamma/\omega$ vanish, the Reynolds number $\rm Re$ being the only parameter left in the Holton-Lindzen-Plumb model.

\subsection{Quasilinear dynamics\label{ssec:QLDyn}}
    We now consider four important approximations for the wave dynamics. Their regime of validity will be discussed \textit{a posteriori} in details in appendix \ref{sec:SelfConsHolton-Lindzen-Plumb}.
    \begin{itemize}
    \item[] \textbf{Quasilinear approximation:} we ignore the nonlinear terms $\nabla\cdot\left(u'\bm{u}'\right)$ and $\nabla\cdot\left(b'\bm{u}'\right)$ in \eqref{eq:WaveEq} and keep the leading order terms in the Taylor expansion of the bottom boundary condition \eqref{eq:BC_cond} with respect to the bottom elevation $h$. This yields no-slip condition for the mean-flow at a flat boundary 
    \begin{equation}\label{eq:BC_mf}
       \overline{u}|_{z=0}=0
    \end{equation}

    \item[] \textbf{Frozen stratification:} we ignore  $\partial_{z}\overline{b}$ in \eqref{eq:WaveEq_2} such that the stratification profile remains linear at all time.
    \item[] \textbf{Hydrostatic approximation:} we consider the hydrostatic balance in place of the vertical momentum conservation in \eqref{eq:WaveEq}.
    \item[] \textbf{Non-viscous wave:} we ignore $\nu\nabla^2 \bm{u}'$ in \eqref{eq:WaveEq_1}, as in the original work of \cite{Plumb1975}, and  we reduce the wave boundary condition to an impermeability condition 
    \begin{equation}\label{eq:BC_Noslip_QL}
        w'|_{z=0}=\partial_t h.
    \end{equation}
    The effect of viscous damping in the domain bulk has been discussed by \cite{Plumb1977}, in particular to discuss laboratory experiments \citep{Plumb1978}. However in both papers the boundary layers associated with the viscous boundary condition are ignored without further comments, which is not always satisfactory \citep[see][]{renaud_venaille_2019}.
    \item[] \textbf{Timescale separation:} we assume that waves evolve on a much faster timescale than the mean flow such that they adjust instantaneously to any change in the mean-flow profile. Therefore, we compute the stationary wavefield assuming a frozen-in-time mean flow. 
    \end{itemize}

    To compute the stationary wavefield under these hypotheses, it is convenient to introduce the streamfunction $\psi'$ such that $(u',w')=(-\partial_z \psi',\partial_x \psi')$ and to decompose the boundary undulation and this streamfunction  on Fourier modes
    \begin{equation}\label{eq:mpwEq}
    \left(h_b\left(x,t\right),\psi'\left(x,z,t\right)\right)=\frac{1}{2}\sum_{n}\left(h_{n},\psi_{n}\left(z\right)\right)\,\ep^{i\left(\omega_n t-k_nx\right)}+\text{c.c.}    
    \end{equation}
    where $\omega_n$, $k_n$ and $h_{n}$ denote the angular frequency, the horizontal wave number and the complex amplitude of each modes and "c.c." stands for "complex conjugate". Then, each mode $\psi_n(z)$ obeys the Taylor-Goldstein equation
    \begin{equation}\label{eq:Taylor-Goldstein}
        \left\{\partial_z^2+\left(\frac{k_n^2N^2}{(\omega_n-k_n\overline{u}(z))^2+\gamma^2}\left(1+i\frac{\gamma_n}{\omega_n-k_n\overline{u}(z)}\right)+\frac{k_n}{\omega_n-k_n\overline{u}(z)}\partial_z^2\overline{u}(z)\right)\right\}\psi_{n}\left(z\right)=0, 
    \end{equation} 
    with bottom boundary condition 
    \begin{equation}\label{eq:LinBC_w}
        \psi_{n}(0)=-\frac{\omega_n h_n}{k_n}.
    \end{equation}
    The upward flux of horizontal momentum (momentum flux hereafter) is constituted of the sum each individual mode contribution
    \begin{equation}\label{eq:momfluxPsi}
        \overline{u'w'}(z)=\sum_{n}\overline{u'_nw'_n}(z),\quad\text{with} \quad \overline{u'_nw'_n}(z)=\frac{k_n}{4}\left(\psi^*_n\partial_z\psi_n-\text{c.c.}\right).
    \end{equation}
   where the averaging operator is defined\footnote{Without the time filtering in this definition, there would be additional cross terms corresponding to high frequency oscillations.} in \eqref{eq:WMF_decomp}.
   
    In \eqref{eq:MeanFlowEq}, the mean-flow is forced by minus the divergence of the momentum flux \eqref{eq:momfluxPsi}. In the next subsection, we derive a closed-form for the momentum flux using additional approximations also discussed in appendix \ref{sec:SelfConsHolton-Lindzen-Plumb}.
    
    \subsection{Momentum flux closure}
    Since the momentum flux $\overline{u'w'}$ is a sum over contributions from independent wave modes (see Eq.\ \eqref{eq:momfluxPsi}), let us start by considering a single mode with amplitude $h$, wavenumber $k$ and angular frequency $\omega$. Indices will be added back later on when considering multiple modes.
    \subsubsection*{Homogeneous case}
\begin{figure}[t!]
    \hfill
    \begin{subfigure}[t]{0.48\textwidth}
        \centering
        \includegraphics[width=\linewidth]{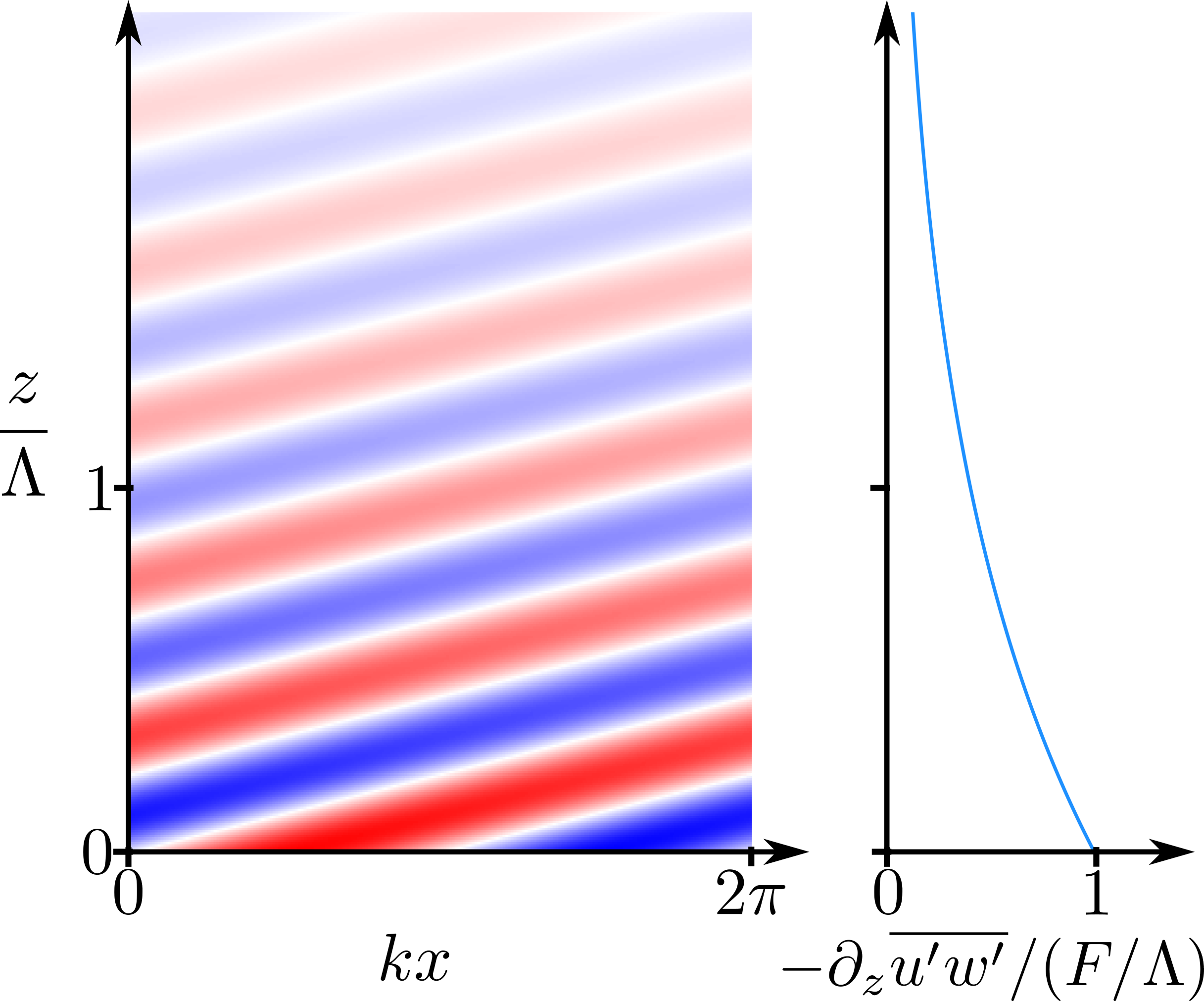}
        \caption{\label{subfig:homStream}Homogeneous streaming}
    \end{subfigure}%
    \hfill
    \begin{subfigure}[t]{0.48\textwidth}
        \centering
        \includegraphics[width=\linewidth]{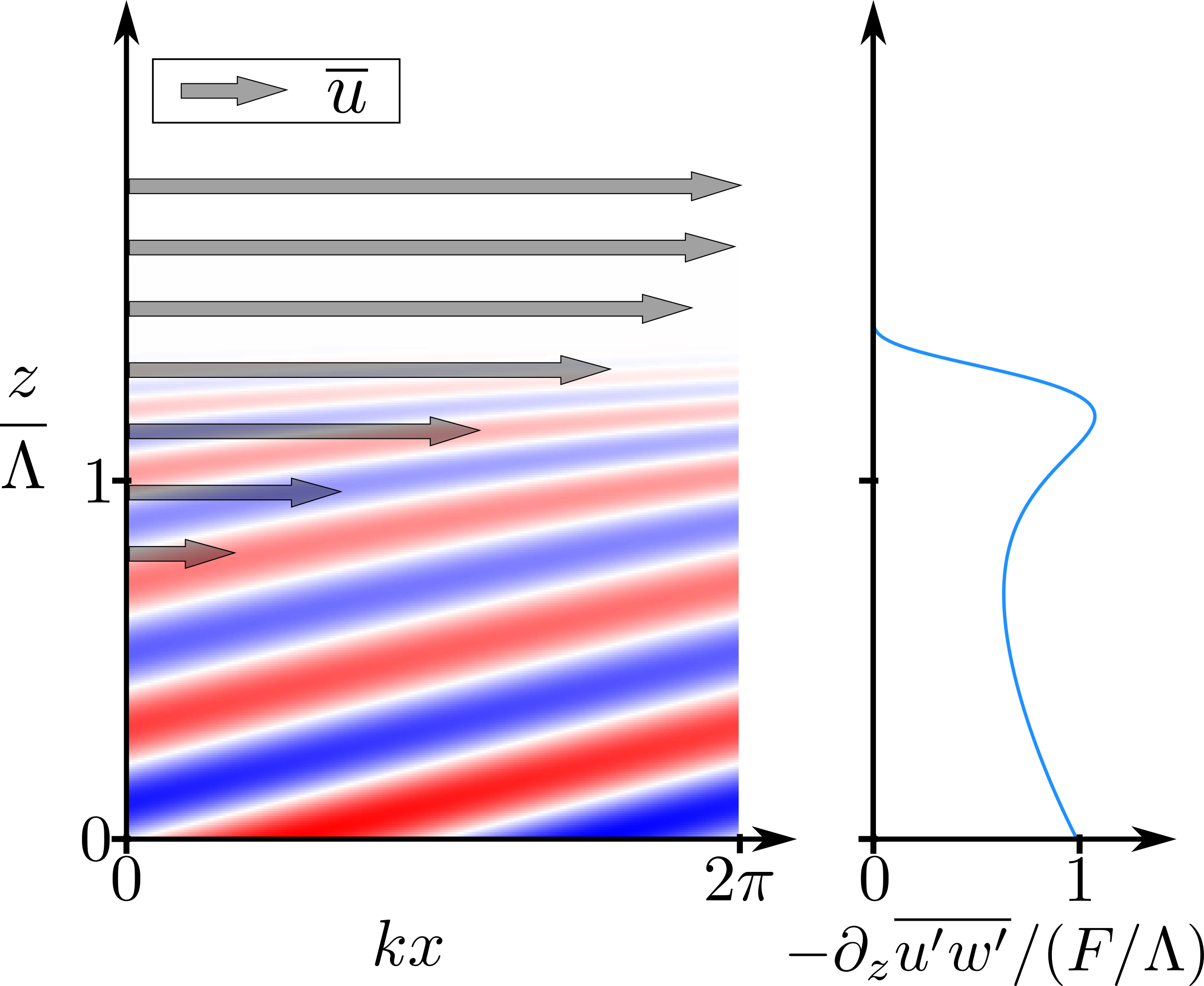}
        \caption{\label{subfig:ShearStream}Shear streaming}
    \end{subfigure}
    \hfill
    \caption{\label{fig:SketchStreaming}\textbf{Wave streaming.} Snapshot of the vertical velocity field associated with a damped monochromatic progressive plane wave propagating within a resting flow (a) or within a shear flow (b) sketched with grey arrows accompanied with a plot of their associated momentum flux divergence vertical profiles.}
    \end{figure}

    Important physical insights can be gained considering first the case without mean flow ($\overline{u}=0$). In this case, the Taylor-Goldstein equation  \eqref{eq:Taylor-Goldstein} is homogeneous and its solution takes the form of a damped vertical oscillation
    \begin{equation}
         \label{eq:VeritcalMode}
        \psi\left(z\right)=\psi(0)  \exp\left\{-imz-\frac{z}{2\Lambda}\right\}
    \end{equation}
where $m\in \mathbb{R}$ is the real part of the vertical wavenumber characterising the oscillation, $\Lambda>0$ is a damping length and $\psi(0)$ is given by \eqref{eq:LinBC_w}. We now make a \textbf{weak damping} approximation, by assuming $\gamma\ll\omega$. 
    Then, injecting ansatz \eqref{eq:VeritcalMode} in \eqref{eq:Taylor-Goldstein}, we find at leading order in $\gamma/\omega$
    \begin{equation}
        m=-\frac{|k|N}{\omega}\quad \text{and}\quad\Lambda=\frac{\omega^2}{\gamma N |k|}\label{eq:mSolut}.
    \end{equation}
Using Eqs. \eqref{eq:LinBC_w} and \eqref{eq:momfluxPsi}, the momentum flux reads at leading order
    \begin{equation}\label{eq:momflux_hom}
        \overline{u'w'}(z)= F \ep^{-z/\Lambda} , \text{ with } F\equiv  \text{sign}(k)\frac{N\omega|h|^2}{2} . 
    \end{equation}
 Figure \ref{subfig:homStream} shows a snapshot of the damped wave vertical velocity field $w'$ and the corresponding momentum flux divergence $-\partial_z\overline{u'w'}$. The latter decays exponentially with height over a scale corresponding to the damping length $\Lambda$.
    
    The damping length $\Lambda$ has an interpretation in terms of the inviscid upward group velocity \citep{Vallis2017}:
    \begin{equation}\label{eq:lambda_cg}
    \Lambda=\frac{w_g}{\gamma}\quad\text{and}\quad w_g=\frac{\partial \omega}{\partial m}=\frac{\omega^2}{|k|N}.
    \end{equation}
    In the presence of a mean flow, the $\omega^2$ dependence in $\Lambda$ will impact dramatically the vertical momentum flux profile, as we now discuss. 

     \subsubsection*{Inhomogeneous case}
     Let us now consider an arbitrary frozen mean flow profile $\overline{u}(z)$. With clear scale separation between the wave and the mean flow, we expect the wave to behave locally as in the homogeneous case, but doppler shifted:
     \begin{equation}
         \hat{\omega}(z)=\omega\left(1-\frac{\overline{u}(z)}{c}\right)
     \end{equation}
     where $c=\omega/k$ the horizontal phase speed. Considering a \textbf{global weak damping limit} with $\gamma\ll\hat{\omega}$ at all $Z$, the local damping length also varies by a factor $\hat{\omega}^2$. The wave is therefore damped more rapidly at heights where the mean flow approaches the horizontal phase speed $c$. Consequently, we expect enhanced streaming in that case. Formally, this behaviour is captured by the leading order terms of a WKB expansion of the wave. Before performing this expansion, let us introduce  dimensionless variables 
    \begin{equation}
        Z=\frac{z}{\Lambda}\quad\text{and}\quad\mathfrak{m}(Z)=-\frac{\omega}{\hat{\omega}(Z)},
    \end{equation}
    where $\Lambda$ is the damping length introduced in \eqref{eq:lambda_cg}, and $\mathfrak{m}$ is a rescaled local vertical wave number. It will also be convenient to introduce a Richardson number 
    \begin{equation}
         \Ri\equiv\left(\frac{N}{c/\Lambda}\right)^2 =\left(\frac{\omega}{\gamma}\right)^2 \label{eq:DEF_Richardson} .
    \end{equation}
    With these notations, the Taylor-Goldstein equation \eqref{eq:Taylor-Goldstein} reads
    \begin{equation}\label{eq:TG_Rescaled}
       \frac{1}{\Ri}\partial_Z^2\psi+\left(\frac{\Ri\,\mathfrak{m}^2}{\Ri+\mathfrak{m}^2}\left(1-i\frac{\mathfrak{m}}{\Ri^{1/2}}\right)-\frac{\mathfrak{m}}{\Ri}\partial_Z^2\frac{1}{\mathfrak{m}}\right)\psi=0, 
    \end{equation} 
    The \textbf{weak damping} limit ($\gamma\ll \omega$) leads to a large Richardson limit ($\Ri \gg 1$), guaranteeing the stability of the mean-flow with respect to Kelvin-Helmholtz instability. The WKB parameter will be given by $\Ri^{-1/2}$; self-consistency of the WKB approach can thus be considered as a consequence of the weak damping limit.
    
    To simplify the Taylor-Goldstein equation, we assume a \textbf{lengths scale separation} 
        \begin{equation}\label{eq:CriticalLayer_WKB_assomp}
        \left|\frac{1}{\mathfrak{m}\,\Ri}\partial_Z^2\frac{1}{\mathfrak{m}}\right|\ll|\frac{\mathfrak{m}}{\Ri^{1/2}}|\ll 1
        \end{equation}
    at all $Z$. At order one in $\mathfrak{m}/\Ri^{1/2}$, the Taylor Goldstein equation reads 
    \begin{equation}\label{eq:TG_Simple}
         \frac{1}{\Ri}\partial_Z^2\psi+\mathfrak{m}^2\left(1-i\frac{\mathfrak{m}}{\Ri^{1/2}}\right)\psi=0. 
    \end{equation} 
    We approximate the solution of this equation by the leading order terms of the WKB expansion
    \begin{equation}\label{eq:WKB_ansatz}
        \psi(Z)=\exp\left\{i\,\Ri^{1/2}\sum_{j=0}^{\infty}\frac{g_j(Z)}{\Ri^{j/2}} \right\},
    \end{equation}
    where the $g_j$ are complex functions of $Z$. Injecting \eqref{eq:WKB_ansatz} in \eqref{eq:TG_Rescaled} and collecting the zeroth order terms yields $g_0(Z)=\pm\int_0^Z\mathfrak{m}(Z')\dd Z'$. Collecting the first order terms leads to an expression for $g_1$ that depends on $g_0$. Keeping the solution that vanishes at infinity, we obtain at this order the WKB expression
    \begin{equation}\label{eq:WKB_Expression}
        \psi(Z)=\psi(0)\left|\frac{\mathfrak{m}(0)}{\mathfrak{m}(Z)}\right|^{1/2}\exp\left\{-i\,\Ri^{1/2}\int_0^Z\mathfrak{m}(Z')\dd Z'-\frac{1}{2}\int_0^Z \mathfrak{m}^2(Z')\dd Z'\right\},
    \end{equation}
    where $\psi(0)$ is determined by the boundary condition \eqref{eq:LinBC_w}.    
    Changing variable back to $z$ and $\overline{u}$, the mean momentum flux reads 
    \begin{equation}
        \overline{u'w'}(z)=F \exp\left\{-\frac{1}{\Lambda}\int_{0}^{z}\frac{\mathrm{d}z'}{\left(1-\overline{u}(z')/c\right)^{2}}\right\}\label{eq:WKB_Reduced_MomFlux},
    \end{equation}
with $F$ defined in Eq.\ (\ref{eq:momflux_hom}). Note that the momentum flux and the horizontal phase speed have the same sign whatever the mode considered ($F\,c>0$). Figure \ref{subfig:ShearStream} shows a snapshot of the wave vertical velocity field $w'$ propagating through a background shear flow $\overline{u}$ represented with grey arrows and the corresponding streaming force $-\partial_z\overline{u'w'}$. The shear background flow enhances the streaming force close below the critical height where $\overline{u}=c$. However, the vertically integrated streaming force does not depend on the background shear profile. A positive background shear flow thus concentrates the streaming force to lower heights. 
   
   \subsubsection*{Closed mean flow equation}
    
   Eq. \eqref{eq:WKB_Reduced_MomFlux} offers a closed form for the momentum flux due to a single-mode. This expression depends solely on the instantaneous mean flow vertical profile $\overline{u}$. It is straightforward to generalise this result to multiple modes indexed by $n$.   The mean flow evolution equation \eqref{eq:MeanFlowEq} becomes then a one-dimensional integrodifferential equation
    \begin{equation}\label{eq:MeanFlow_General}
        \partial_t\overline{u}-\nu\partial_z^2\overline{u}=-\partial_z\left(\sum_n F_n\exp\left\{-\frac{1}{\Lambda_n}\int_0^z\frac{\mathrm{d}z'}{(1-\overline{u}(z')/c_n)^2}\right\}\right),
    \end{equation}
    where $F_n$, $\Lambda_n$, and $c_n$  are the bottom momentum flux, the dissipation length and the horizontal phase speed of the $n$-th mode such that $F_nc_n\leq 0$. 
    
    Equation \eqref{eq:MeanFlow_General} is the model derived originally in \citep{Plumb1975,Plumb1977}, building on the physical insights from Holton and Lindzen theory of quasi-biennial oscillation \citep{Holton1972}. 
    We followed most of the steps of the original derivation by \cite{Plumb1975}, collected and rephrased all the required approximations in a systematic manner. 
    Furthermore, we will check \textit{a posteriori} in Appendix \ref{sec:SelfConsHolton-Lindzen-Plumb} that the set of approximations made along the way are self-consistent. For that purpose, we will need a deeper characterisation of the mean flow evolution through equation \eqref{eq:MeanFlow_General} which we now tackle.
\section{Solutions of the Holton-Lindzen-Model in simple cases \label{sec:MF-Dyn}}

This section investigates the dynamics of the mean-flow generated and steered by the streaming of damped internal waves propagating from below, using Holton-Lindzen-Plumb model derived in previous section \ref{sec:Holton-Lindzen-Plumbmodel}. The resulting uni-dimensional integrodifferential equation \eqref{eq:MeanFlow_General} is easily solved numerically over long timescales using a standard finite difference approach (see \cite{RenaudPHD2018}, appendix A for more details). In a first subsection, we consider the particular case of a single wave streaming in the which the mean-flow ultimately reach a steady state. The addition of a second counter-propagating wave, considered in a second subsection, allows for the mean-flow stationary state to bifurcate from a steady regime to a limit cycle when the wave-streaming force is increased. Both the case of a single wave and two counter-propagative waves were discussed in the original work of Plumb \citep{Plumb1977}. Our contribution is to give an analytical solution to the steady-state solution in the single wave case, which allows to interpret the numerical results in the two counter-propagating wave case and to obtain \textit{a posteriori} scaling laws on typical length scales and velocities. These scalings are used in appendix \ref{sec:SelfConsHolton-Lindzen-Plumb} to check the self-consistency of the model derived in the previous section under a number of hypotheses. 

\subsection{Single wave streaming}

Let us consider the mean-flow evolution induced by a single rightward propagating wave with amplitude $h$.  The associated closed mean flow evolution equation reads
\begin{equation}\label{eq:MeanFlow_single}
        \partial_t\overline{u}-\nu\partial_z^2\overline{u}=-F\partial_z\left(\exp\left\{-\frac{1}{\Lambda}\int_0^z\frac{\mathrm{d}z'}{(1-\overline{u}(z')/c)^2}\right\}\right),
\end{equation}
where the horizontal phase speed $c$ and the bottom upward momentum flux $F$ defined in Eq.\  (\ref{eq:momflux_hom}) of the wave are both positive. Natural characteristic length and velocity are provided by the damping length $\Lambda$ and the phase speed $c$. A characteristic timescale for the streaming is defined by
\begin{equation}
    \tau=\frac{c\Lambda}{F}. \label{eq:tauDEF}
\end{equation}
This time scale corresponds to the period scaling given in \cite{Vallis2017}. Considering the dimensionless variables
\begin{equation}\label{eq:DimLess_Var}
Z=\frac{z}{\Lambda},\quad T=\frac{t}{\tau}\quad\text{and} \quad U=\frac{\overline{u}}{c},
\end{equation}
the mean flow equation now  reads 
\begin{equation}\label{eq:MeanFlow_single_scaled}
        \partial_TU-\frac{1}{ \Rey }\partial_Z^2U=-\partial_Z\left(\exp\left\{-\int_0^Z\frac{\mathrm{d}Z'}{(1-U)^2}\right\}\right),
\end{equation}
where 
\begin{equation}\label{eq:Reynolds}
     \Rey =\frac{\Lambda F}{\nu c},
\end{equation}
is the Reynolds number introduced previously equation \ref{eq:NondimensionalNumbers}. This is the single control parameter of the model. Qualitatively, it compares the strength of the wave forcing to the mean viscous stress. Equation \eqref{eq:MeanFlow_single_scaled} is coupled with the boundary conditions $U|_{Z=0}=0$ and $\partial_ZU|_{Z\to\infty}=0$. In numerical application, an upper flat boundary is located at $z=1.5\Lambda$ with a free slip boundary condition, which induces finite size effect but does not change the qualitative behaviour of the system. 

\subsubsection*{Downward propagation}
  \begin{figure}
        \centering
        \includegraphics[width=0.45\linewidth]{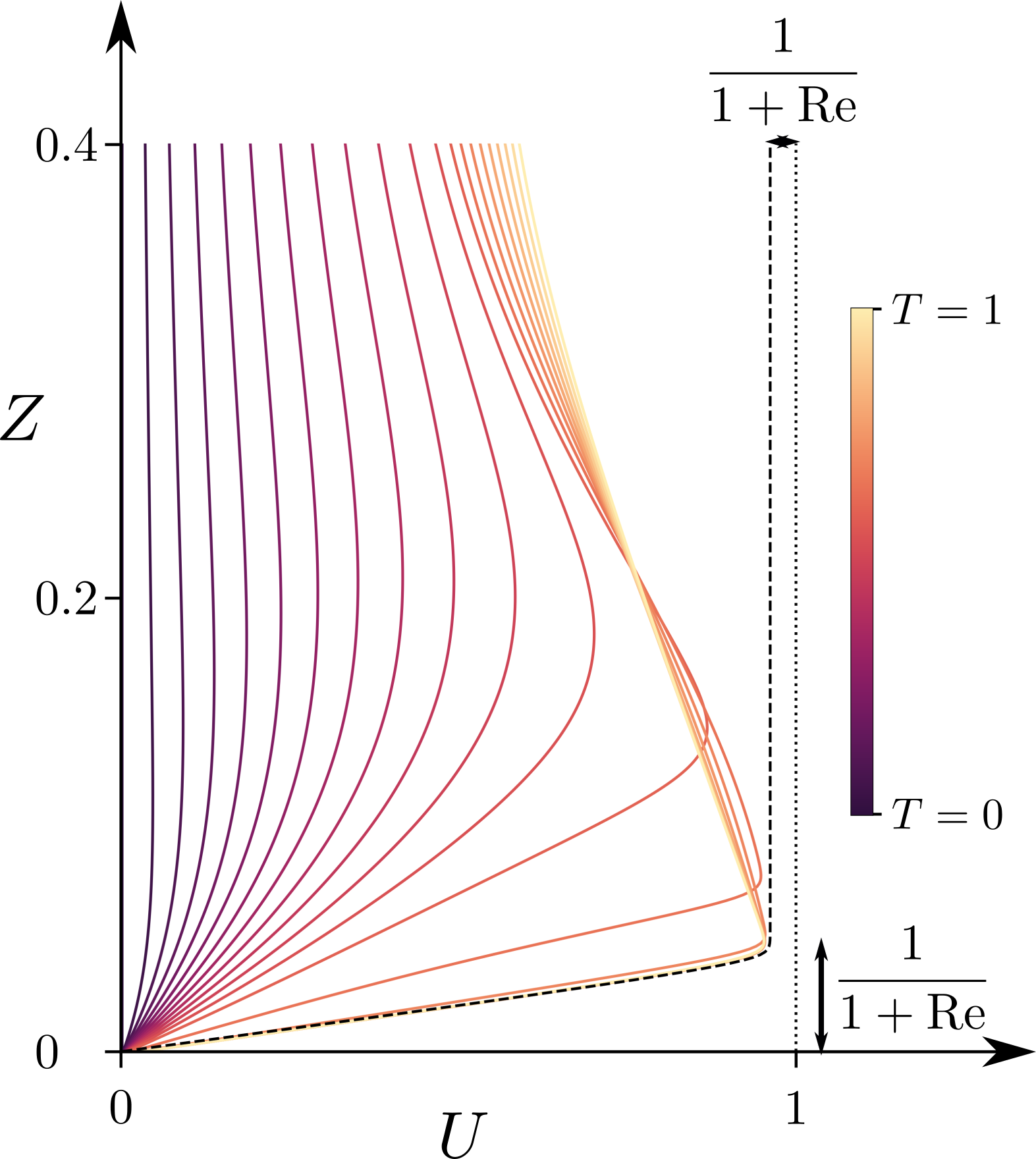}
        \caption{\textbf{Single wave streaming and downward mean flow propagation.} Snapshots of the mean flow vertical profile are shown obtained by direct numerical resolution of Eq. \eqref{eq:MeanFlow_single_scaled} using $ \Rey =25$. The stationary solution given by Eq. \eqref{eq:SS_profile} is shown in dashed.}\label{fig:Descend}
    \end{figure}
We integrate \eqref{eq:MeanFlow_single_scaled} numerically starting from rest for $ \Rey =25$. The evolution of the mean-flow profile is shown in figure \ref{fig:Descend}. The streaming force on the right-hand side of \eqref{eq:MeanFlow_single_scaled} is everywhere positive and is henceforth forcing a rightward mean flow. 

A characteristic dynamical feature arises when $U$ reaches order $1$ values: the streaming force profile get confined to lower levels leading to a downward propagation of the mean flow (see figure \ref{fig:Descend} for $T>0.5$). For $T>1$, the streaming force is locally balanced by the viscous stress. There remains only the slow viscous diffusion of momentum above the critical layer until a steady state is reached at $T\to\infty$.

Downward propagation of the QBO phase through a streaming mechanism was first noticed by \cite{Lindzen1968} and has been a  major achievement of Holton-Lindzen-Plumb theory. Plumb showed numerical simulations of the relaxation towards a steady-state \citep{Plumb1977},  as in figure \ref{fig:Descend}. Recent laboratory experiments have also described this single-wave streaming phenomenon \citep{Semin2016}, in which case damping is dominated by viscosity. To our knowledge, there has been no analytical description of the limiting steady state in these different regimes. 

\subsubsection*{Steady state}
The steady state $U_{\infty}(Z)$ satisfies 
\begin{equation}\label{eq:SS_eq}
    \partial_Z^2U_{\infty}=\Rey\,\partial_Z\left(\exp\left\{-\int_0^Z\frac{\mathrm{d}Z'}{(1-U_{\infty})^2}\right\}\right).
\end{equation}

Using the bottom no-slip boundary condition, and free-slip condition at infinity, this equation admits the analytical solution 
\begin{equation}\label{eq:SS_profile}
    U_{\infty}(Z)=\frac{ \Rey -W( \Rey \;\ep^{ \Rey -(1+ \Rey )^2Z})}{1+ \Rey },
\end{equation}
where $W$ denotes the Lambert-W function ($y=w \ep^w\;\Longleftrightarrow \;w=W(y)>-1$). A detailed derivation is provided in appendix \ref{sec:SSComp}. The mean flow profile \eqref{eq:SS_profile} is shown in dashed in figure \ref{fig:Descend} for $ \Rey =25$. In the low Reynolds number limit, we recover steady solution of the homogeneous problem $U_{\infty}(Z)=  \Rey\,(1-\ep^{-z})$.

Expression \eqref{eq:SS_profile} is useful to estimate how close the mean flow approaches the critical value $U=1$ with 
\begin{equation}\label{eq:umax}
    U_{\rm max}=\lim_{Z\to\infty}U_{\infty}(Z)=\frac{ \Rey }{1+ \Rey }.
\end{equation}
Moreover, in the large Reynolds number limit the steady mean flow takes the form 
\begin{equation}\label{eq:SS_Prof_LargeRey}
U_{\infty}(Z)=\min\{1, \Rey \, Z\}
\end{equation}
with a characteristic scale of $ \Rey ^{-1}$ for the bottom shear. These estimates for the steady flow profile are useful to check consistency of the derivation for Holton-Lindzen-Plumb model (see appendix \ref{sec:SelfConsHolton-Lindzen-Plumb}). 

\subsection{\label{ssec:CounterWaveStream} Symmetric counterpropagating waves streaming} 

\begin{figure}
        \centering
        \includegraphics[width=0.6\textwidth]{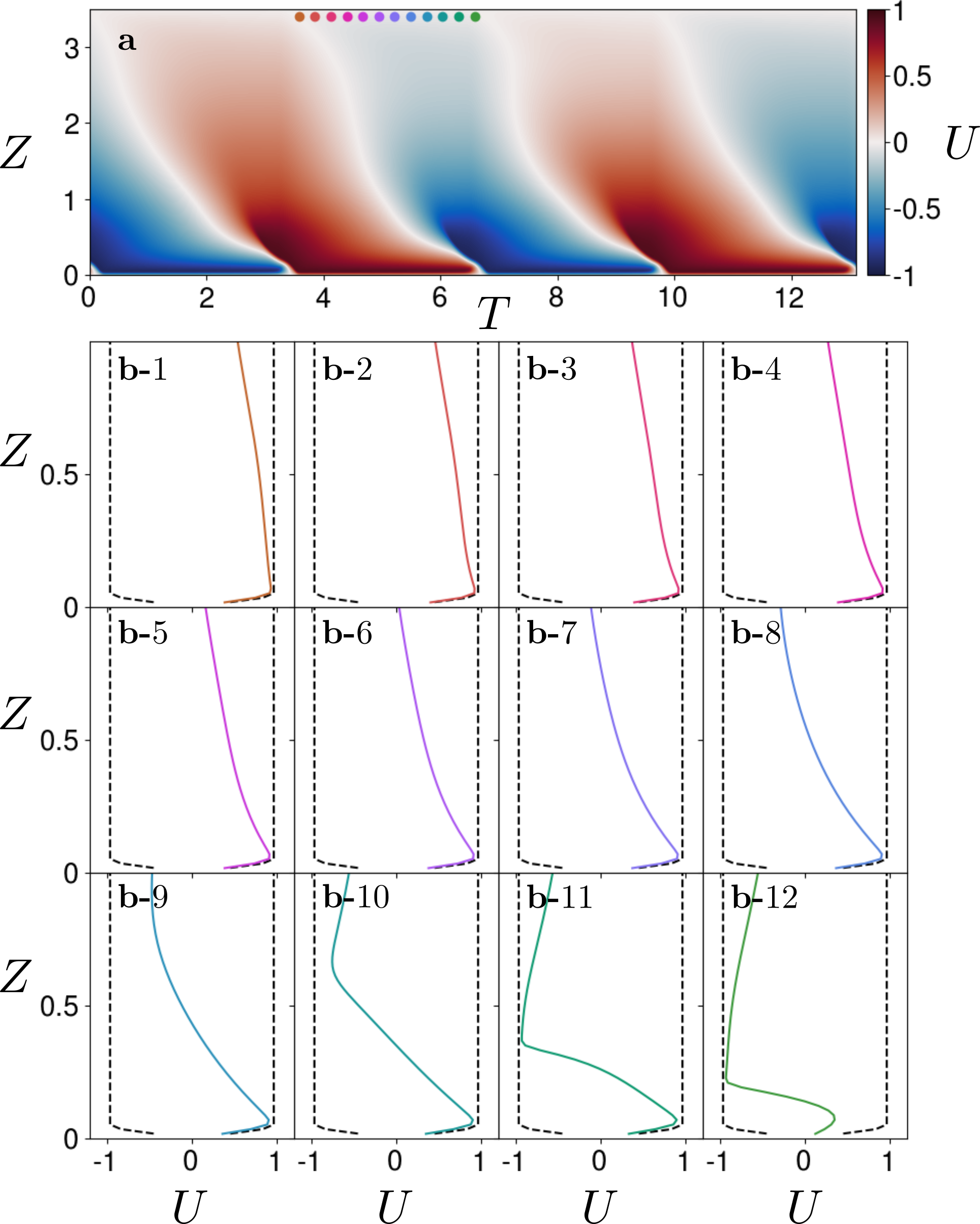}
        \caption{\textbf{Mean flow reversals. a.} Time-height section of two periods of the mean flow limit cycle obtained by numerically integrating Eq. \eqref{eq:Holton-Lindzen-Plumb} with $ \Rey =25$. \textbf{b-1} to \textbf{b-12}. Snapshots of the mean flow vertical profile evenly spaced in time within half a period of the limit cycle. The dashed lines represent the mean-flow steady-state profiles associated with the streaming of the two forcing waves taken individually.}
        \label{fig:ModelOut} 
\end{figure}

As noticed by \cite{Plumb1977}, the simplest setting leading to spontaneous mean flow reversals corresponds to the mean flow evolution \eqref{eq:MeanFlowEq} driven by two counterpropagating waves with equal amplitude and frequencies, but opposite wavenumbers. Then, using the dimensionless variables introduced Eq. (\ref{eq:DimLess_Var}), the closed mean flow evolution equation reads
    \begin{equation}\label{eq:Holton-Lindzen-Plumb}
\partial_{T}U- \Rey ^{-1}\partial_Z^2U=-\partial_Z\left(\exp\left\{-\int_0^Z\frac{\mathrm{d}Z'}{\left(1-U\right)^2}\right\}-\exp\left\{-\int_0^Z\frac{\mathrm{d}Z'}{\left(1+U\right)^2}\right\}\right),
    \end{equation}
complemented by the boundary conditions $U|_{Z=0}=0$ and $\partial_ZU|_{Z\to\infty}=0$. In this symmetric case, the rest state $U=0$ is a natural fix point to the dynamics which is always stable at low Reynolds numbers. Above a critical Reynolds number $\Rey _c\approx 4.15$, the rest state becomes unstable and the system reaches an oscillating state which presents the salient features of the quasi-biennial oscillation \citep{Plumb1977}. Numerical computations of these oscillations in the case $ \Rey =25$ are shown in figure \ref{fig:ModelOut} . Snapshot of the mean-flow vertical profile evenly spaced within half a period of the cycle is shown from \ref{fig:ModelOut}b-1 to \ref{fig:ModelOut}b-12, together with the steady states of single-wave streaming: From b-1 to b-9, the rightward streaming force is balanced with the viscous stress at the bottom and the leftward streaming force is pushing the mean-flow leftward above the critical layer. The bottom part of the flow is close to the steady-state of the single-wave configuration. This goes on until a second critical layer is created in the upper part of the flow (see b-10) which then propagates downward (see b-10 to b-12). At some point, the viscous stress coming from the shear between the two critical layers becomes strong enough that it takes over the rightward forcing: the bottom mean-flow reverses (see b-12). We then end-up in configuration symmetric to b-1 and the second half of the cycle appends following the same scheme. 

\section{Bifurcation diagrams of the Holton-Lindzen-Plumb model\label{sec:BifDiagram}}

In this section, we analyse the symmetric Holton-Lindzen-Plumb equation \eqref{eq:Holton-Lindzen-Plumb} with the lens of dynamical system theory, building on \citep{yoden1988new,semin2018nonlinear}. We first take a close look at the first bifurcation, namely the instability of the rest state. We then describe the quasi-periodic routes to chaos that were found in \cite{kim2001gravity,renaud2019periodicity}. Our contribution is to offer a comparison between no-slip and free-slip boundary condition at the bottom. The free-slip analysis  detailed in appendix \ref{sec:free-slip} is new. The comparison brings to light significant differences between these two boundary conditions, even if the global structure of the bifurcation diagram is left unchanged. 

\subsection{Instability of a rest state}

\begin{figure}
    \centering
    \includegraphics[width=0.65\linewidth]{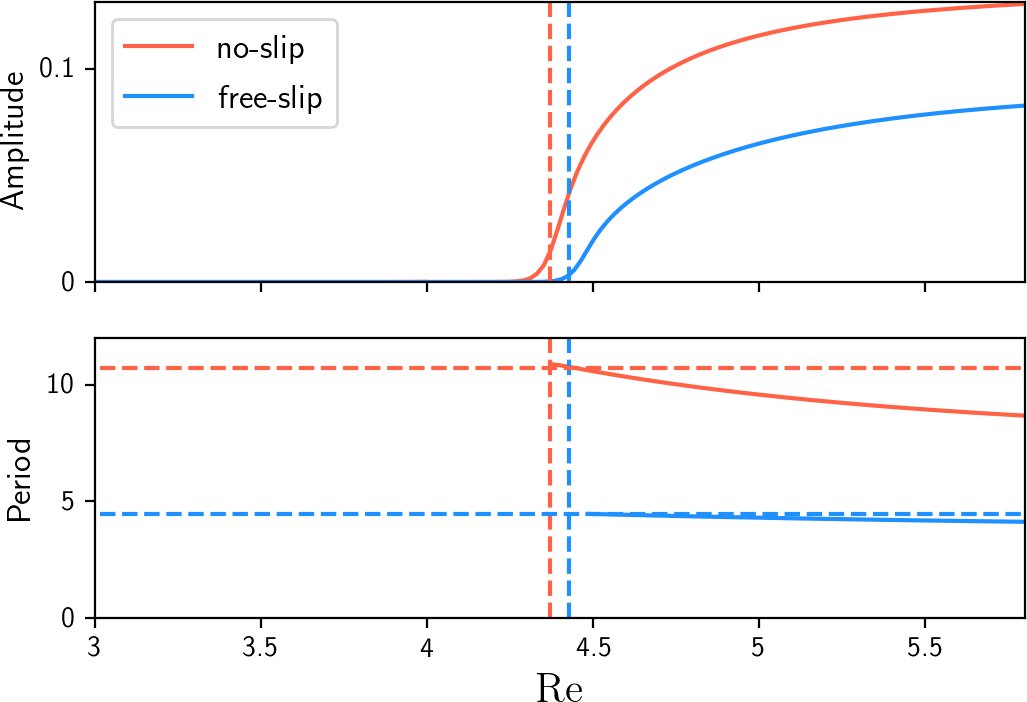}
    \caption{\textbf{Instability of the rest state in the Holton-Lindzen-Plumb model.} Plot of the amplitude of the limit cycle at $Z=3$ (top panel, solid lines) and, if applicable, its period (bottom panel, solid lines) as a function of the Reynolds number. This result is obtained by solving Eq. \eqref{eq:Holton-Lindzen-Plumb} numerically using a no-slip (red) or a free-slip (blue) boundary condition, starting above the threshold and decreasing the Reynolds number slowly. The numerical upper boundary was set at $Z=4$ and  $500$ vertical grid points were used.  The dashed lines show the theoretical critical values computed in appendices \eqref{sec:FirstBif} and \eqref{sec:free-slip}.}
    \label{fig:FirstBiff}
\end{figure}

When the Reynolds is sufficiently small, the rest state is a stable fixed point of the Holton-Lindzen-Plumb equation \eqref{eq:Holton-Lindzen-Plumb}. For sufficiently large Reynolds number it becomes unstable. The basic mechanism underlying this instability was qualitatively understood by \cite{Holton1972}. 
Building on their interpretation, and using a linearized version of   \ref{eq:Holton-Lindzen-Plumb}, Plumb provided a quantitative analysis of this instability mechanism: close to a rest state with $U\ll 1 $ and  $ \int^z U \mathrm{d} z' \ll 1 $. In this limit, the forcing term on the right-hand side can be simplified, leading to 
\begin{equation}\label{eq:Holton-Lindzen-Plumbsimplified}
    \partial_{T}U- \Rey ^{-1}\partial_Z^2U= 4 \left( U-\int^z U \mathrm{d} z' \right)\ep^{-Z}\ .
\end{equation}
The first terms of the r.h.s. corresponds to positive feedback between the mean flow and the enhanced streaming induced by the wave with a phase speed having locally the same sign as the mean flow. The second term involves the vertically integrated velocity field and can be interpreted as a shielding term. Assuming the velocity field is initially positive, the shielding takes over the enhanced streaming at hight altitude, changing the sign of the r.h.s.:   the wave propagating in the same direction as the mean flow has been efficiently damped so that the streaming becomes dominated by the contribution from the counter-propagating wave.
\cite{Plumb1977} showed that the rest state is always unstable in the limit of infinite Reynolds numbers. \cite{yoden1988new} studied numerically the details of the instability, including a discussion on the effect of asymmetric wave forcing, and a two-layer version of \eqref{eq:Holton-Lindzen-Plumb} . A semi-analytical computation of the marginal instability curve is obtained by \cite{semin2018nonlinear} in a case where the wave attenuation is dominated by viscosity with a no-slip bottom boundary condition for the mean horizontal flow. They found good agreement with experimental data. We reproduce in appendix \eqref{sec:FirstBif} their computation of the instability threshold, focusing on the Holton-Lindzen-Plumb model \eqref{eq:Holton-Lindzen-Plumb} where wave attenuation is dominated by Newtonian cooling. The critical Reynolds number and the period of the oscillation at the threshold are found to the solution of a transcendental equation (see eq.\ \eqref{eq:RootEigen}). A numerical resolution yields the critical Reynolds number  $Re_{c}\approx 4.37$ and the critical period  $T_c\approx10.7$. Figure \eqref{fig:FirstBiff} compares these prediction to direct simulation of the nonlinear Holton-Lindzen-Plumb equation \eqref{eq:Holton-Lindzen-Plumb}.

\subsubsection*{Nonlinear saturation close to the first bifurcation}

Yoden and Holton found numerically that the Hopf bifurcation associated with instability of rest state in \eqref{eq:Holton-Lindzen-Plumb} is supercritical. 
The nonlinear saturation of the instability close the first bifurcation threshold was addressed by \cite{semin2018nonlinear} by using multiple-scale analysis using viscous damping and linear friction in the momentum equation. Such terms are relevant to model the effect of lateral walls in laboratory experiments or to account for radiations of waves outside the equatorial region in geophysical flows. They found that the Hopf bifurcation becomes sub-critical when linear damping parameter exceeds a threshold, in good agreement with laboratory experiments.

\subsubsection*{Effect of breaking mirror symmetry in zonal direction}
So far, we have discussed bifurcation diagrams for boundary conditions admitting a mirror symmetry in the x-direction. This symmetry can be broken by increasing the amplitude of one of the two counter-
propagative
 waves at the boundary. It is also possible to break mirror symmetry in problem just by changing properties of wave propagation in the domain bulk, for instance by considering the effect of rotation, as in the original work of \cite{Holton1972}.

As noticed by  \cite{semin2018nonlinear}, the phenomenology of supercritical/subcritical Hopf bifurcation does not change when this mirror symmetry is broken, as the normal form for the bifurcation just depends on the assumption of translational invariance in time for the mean vertical wind profile. The only consequence of breaking mirror symmetry is that the low Reynolds stable state is no more a rest state, and the oscillatory solution presents asymmetries between eastward and westward phases \citep{Holton1972,Plumb1977,yoden1988new}.

\subsection{The case of free-slip boundary conditions}
Up to now, we have only considered the case of no-slip boundary conditions, as in the vast majority of work dealing with Holton-Plumb-Lindzen model. As far as applications to the atmosphere are concerned, the choice of boundary conditions that would mimic the effect of the tropopause to the stratosphere is not obvious. It is thus natural to ask whether the results obtained with no-slip boundary conditions are robust to other choices. We discuss in appendix \ref{sec:free-slip} the case of free-slip conditions at the bottom. 

The main results derived in this appendix are threefold.  First, in the single wave streaming case, we find that the mean flow does not reach a steady-state as it grows past the singular value $U=1$ at a finite time. Second, when forcing with two counter-propagating waves of equal amplitude, we find that the system undergoes a Hopf bifurcation as in the no-slip case considered in \citep{semin2018nonlinear}. While the critical Reynolds number is close to the no-slip value, we find reversals that are roughly two times faster (see fig. \eqref{fig:FirstBiff}). 
Third, we find that the Holton-Lindzen-Plumb model with free-slip boundary conditions does not admit any self-consistent regime: the contribution from the wave boundary layers in the Reynolds-stress tensor can not be dismissed, contrary to the no-slip case.

\subsection{Secondary bifurcations and quasi-periodic route to chaos}

\begin{figure}
    \centering
    \includegraphics[width=0.45\linewidth]{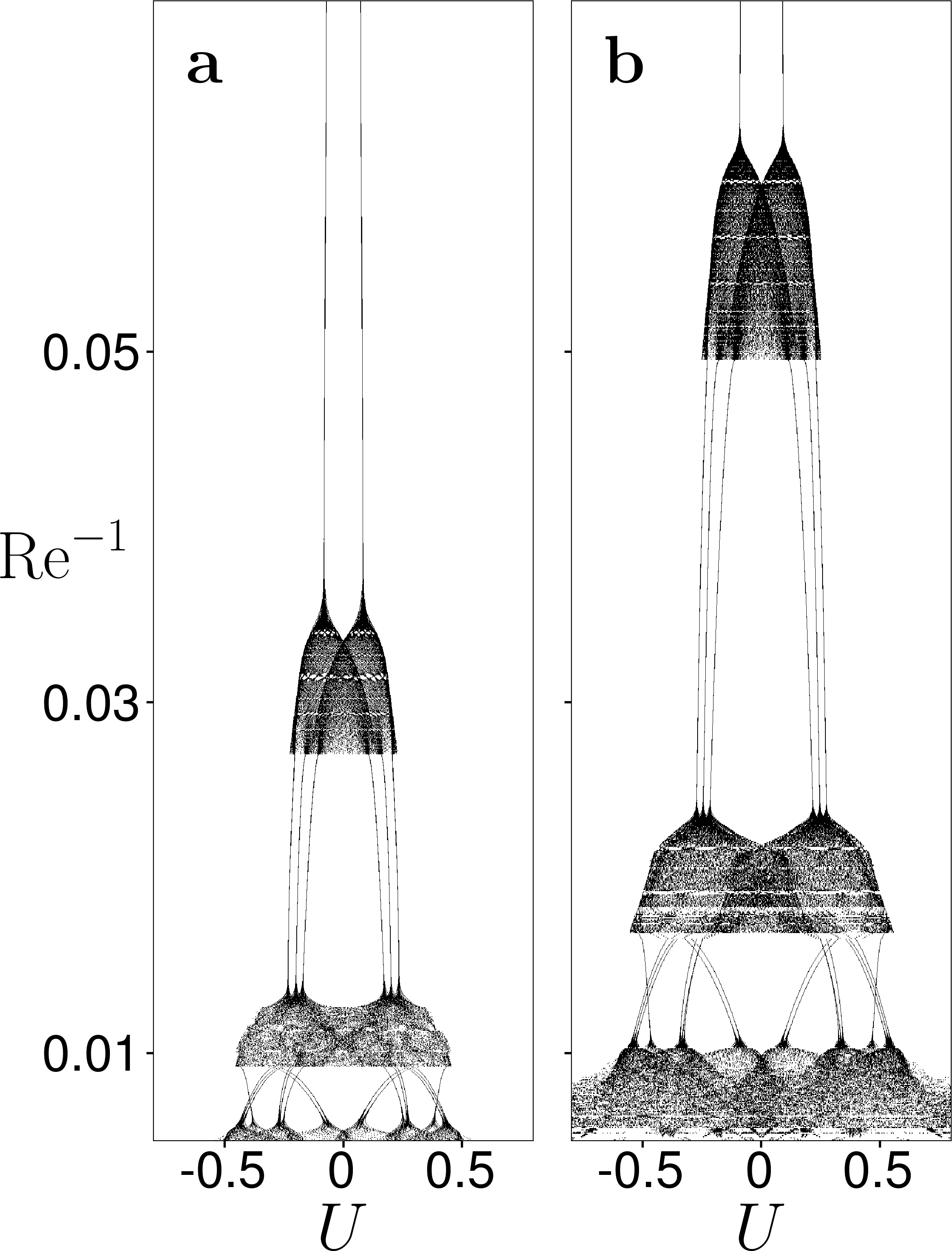}
    \caption{\label{fig:QPR_to_Chaos}\textbf{Quasiperiodic route to chaos in the Holton-Lindzen-Plumb model.} Bifurcation diagrams are shown, obtained for each value of $\Rey^{-1}$ by considering the value of $U$ at two different heights $Z_1$ and $Z_2$, and then by plotting $U(Z_2)$ when $U(Z_1)=0$. Here $Z_1=0.1$ and $Z_2=3$. Panel \textbf{a} shows results obtained by integrating the Holton-Lindzen-Plumb model \eqref{eq:Holton-Lindzen-Plumb} using a no-slip boundary condition at the bottom while panel \textbf{b} shows results associated with the free-slip boundary condition (see appendix \ref{sec:free-slip}).} 
\end{figure}

The full bifurcation diagrams for Holton-Lindzen-Plumb model is plotted figure \ref{fig:QPR_to_Chaos}, both in the no-slip and in the free-slip case, following the procedure of \citep{renaud2019periodicity}. The figure is obtained after many numerical integrations of the model for different values of the control parameter $\mathrm{Re}$, assuming that there is a single attractor for each parameter (as checked numerically by varying the initial condition). We recover in both case the quasiperiodic route to chaos described in \cite{kim2001gravity} and \cite{renaud_venaille_2019}. Starting from a stable rest state and increasing the parameter $\mathrm{Re}$, each bifurcation is associated with shallower mean-flow reversals embedded in slower and deeper oscillations. With respect to the no-slip case, the free-slip boundary condition facilitates the transition to chaos by lowering the successive bifurcations' thresholds significantly.

\section{Conclusion}

In this work, we have revisited the derivation of Holton-Lindzen-Plumb model. By keeping track of all the approximations made and checking their validity a posteriori, we have been able to show the existence of a distinguished limit for which the derivation of the model with two symmetric waves and no-slip bottom boundary condition is self-consistent whatever the value of the control parameter $\mathrm{Re}$. This suggests that the quasiperiodic route to chaos reported in \cite{kim2001gravity} and \cite{renaud_venaille_2019} is an intrinsic property of the original set of Boussinesq equations rather than an artefact of the reduced model. At large Reynolds number, the proof relies on a novel analytical expression of the steady state mean flow associated with each individual wave. We also took a look at the free-slip problem which is rarely consider in the literature. The dynamics presents noticeable differences with respect to the no-slip case: the oscillation periods are about twice slower, and the secondary bifurcation  occurs at much lower Reynolds numbers.

Three-dimensional effects and rotation have been left aside in this study. On the one hand, recent theoretical and experimental work in the non-rotating case and in the $f$-plane case have revealed important modifications of wave-mean flow interactions in the presence of a transverse (meridional), with for instance the generation of vertical vorticity when wave generation varies in the meridional direction \citep{InstabIGW2018}. It will be interesting to investigate these 3D features in the context of QBO-like phenomena. On the other hand, simplified model aiming at describing the meridional extent of the QBO on the equatorial beta plane have been proposed thirty years ago \citep{plumb1982equatorial,plumb1982model,dunkerton1985two}, building on \cite{lindzen1971equatorial} and \cite{Holton1972}. The existence of self-consistent QBO-like models in this beta plane case remains to our knowledge an open question, together with a full description of their bifurcation diagram.

\appendix
\section{Self-consistency of the Holton-Lindzen-Plumb model\label{sec:SelfConsHolton-Lindzen-Plumb}}

We discuss in this appendix the possibility of a self-consistent regime for the Holton-Lindzen-Plumb model considering the simplest case of two symmetric counterpropagating waves. We recall that the control parameter for the bifurcation is the Reynolds number $\Rey= F \Lambda/\nu c = \omega^2 h^2/ (2\gamma \nu)$. To explore the possibility of a distinguished limit leading to the Holton-Lindzen-Plumb model, we have to translate the different hypothesis made for the derivation of the Holton-Lindzen-Plumb model into constraints on the dimensionless numbers. We consider two limiting cases: (i) the system is close to the bifurcation with $\Rey \sim \Rey_c$ such that the mean flow oscillations are weak, (ii) the forcing is large with $\Rey \gg 1$ such that the mean flow approaches the critical layers. 
We look in both cases for a distinguished limit with parameters organised as follows
\begin{equation}
h k = \epsilon ,\quad \frac{\omega}{N} = O(\epsilon^\alpha),\quad \frac{\gamma}{\omega} = O(\epsilon^\beta),\quad\text{and}\quad \Rey= O(\epsilon^{-\delta})  \label{eq:distinguish}
\end{equation}
with $\epsilon\ll1$. In the following we look for triplets of exponents $(\alpha, \beta, \delta)$ consistent with the approximations leading to the Holton-Lindzen-Plumb model.

\subsection{Distinguished limit close to the first bifurcation}

Let us assume that the system is in the QBO-regime, close to the bifurcation threshold, so that $|\overline{u}|\ll c$ everywhere. In other words, the amplitude of the limit cycle remains sufficiently small to avoid critical layers, and vertical variations of $\overline{u}$ are characterised by the damping length $\Lambda$ defined in  Eq. (\ref{eq:mSolut}). We list below the different hypotheses leading to the Holton-Lindzen-Plumb model close to the threshold:

\begin{enumerate}

    \item \textit{System close to the first bifurcation}. The bifurcation threshold occurs for $ \Rey=O(1)$. Assuming that the system is close to this threshold leads therefore to the condition $\delta=0$.
    
    \item \textit{Inviscid wavefield in the bulk.} We ignored the effect of viscosity on the wavefield in the domain bulk, where it is primarily damped by radiative cooling. To neglect the contribution of viscosity we must have $\nu \nabla^2/\gamma\sim \nu m^2/\gamma\ll1$ yielding the condition $ \Rey\, (\gamma/\omega)^2 \gg (hk)^2 (N/\omega)^2$ and thus $2-\delta>2\alpha+2\beta$. 
    
    \item \textit{Weak dissipation limit and WKB parameter.} Throughout this study, we simplified the dispersion relation of internal gravity waves assuming $\gamma\ll \omega$. This implied that the vertical wavelength $1/m$ is much smaller than the attenuation length scale $\Lambda$. The parameter $\gamma/\omega$ also corresponds to the WKB parameter $\Ri^{-1/2}$ used to simplify the computation of the wavefield during the derivation of the model. Thus, the weak dissipation limit guarantees the validity of the WKB approximation. The condition $\gamma/\omega\ll 1$ corresponds to $\beta>0$.

    \item  \textit{Hydrostatic balance}. The condition for hydrostatic balance is satisfied for small (vertical to horizontal) aspect ratio $k/m$, which guarantees that $|\partial_t w'|/|b'|\ll 1 $. Using the dispersion relation  in Eq. (\ref{eq:mSolut}) we obtain the condition $\omega/N \ll 1$ and hereby $\alpha >0$.

    \item \textit{No boundary streaming.} Even if viscosity can safely be neglected in the bulk to compute the wave field, it induces the presence of boundary layers close to the wall. We have neglected streaming induced by these boundary layers. It is hard to justify this hypothesis in the general case  \citep{renaud_venaille_2019}. In the case of a standing wave forcing with no-slip boundary condition,  streaming induced by each of the two counter-propagative waves cancels out if there is no mean flow.  Neglecting boundary streaming in the no-slip case can be justified if the wave boundary layer thickness $\sqrt{\nu/\omega}$ is much smaller than typical length scale for mean flow variations along the vertical. Close to the bifurcation, this length scale is given by the attenuation length scale $\Lambda$. Therefore, we must have $\sqrt{\nu/\omega}\ll \Lambda$, which gives $\Rey^{-1} (hk)^2 (N/\omega)^2 (\gamma/\omega)\ll 1$, and consequently $2+\delta-2\alpha+\beta>0$.
    
    \item \textit{Frozen-in-time stratification.} We assumed the stratification profile to be dominated at any time by the initial one, such that $\partial_z\overline{b}\ll N^2$. The order of magnitude for the mean buoyancy  $\overline{b}$ is estimated from the the steady averaged buoyancy equation \eqref{eq:BoussModel_2}, using  $\overline{b}\sim \partial_z\overline{w'b'}/\gamma$. Taking again the attenuation length scale $\Lambda=\omega^2/(N\gamma k)$ as a  characteristic vertical scale, and using $\overline{w'b'} \sim h^2 \omega  N^2 $  we have $\partial_z\overline{b}/N^2\sim\overline{w'b'}/(\gamma \Lambda^2 N^2) \sim (\gamma/\omega)^2 (N/\omega)^2 (h k)^2$. Therefore the  frozen-in-time stratification condition is fulfilled if $ (\gamma/\omega)^2 (N/\omega)^2 (h k)^2 \ll 1$, corresponding to  $2-2\alpha+\beta>0$. 
    
    \item \textit{Quasi-linear approximation.} We assumed the waves to be linear. This condition is fulfilled if the nonlinear terms are small compared the linear ones in all the equations. Assuming that the smallest linear term is the one involving radiative cooling (in agreement with the weak damping limit above), the linearity condition is satisfied when $|\mathbf{u}'\cdot\nabla|\ll \gamma$. Therefore, we must have $(k h) (N/\omega)(\omega/\gamma)\ll1$ and hereby $\alpha+\beta<1$.
    \item \textit{Time scale separation between waves and mean flows.} The typical  adjustment time for the wavefield around a frozen-in-time mean flow can be estimated as the propagation time for a wave packet over the attenuation length scale $\Lambda$ with vertical group velocity $c_g \sim \omega^2/(kN)$.  We assumed that this adjustment time is much smaller than the typical time for mean flow reversals estimated in Eq. (\ref{eq:tauDEF}) as $\tau \sim c\Lambda /\overline{u'w'}|_0 \sim  \Lambda /(h^2 k N)$. This time scale separation hypothesis is satisfied when 
    This leads to the condition  $(h k)^2 (N/\omega)^2  \ll\tau$, and hence $\alpha<1$. 
\end{enumerate}
All together, the scaling conditions reduce to $4$ with $\alpha>0$, $\beta>0$, $\delta=0$ and $1-\alpha-\beta>0$. A distinguished limit exists for instance with $(\alpha,\beta,\delta)=(1/4,1/4,0)$. 
Note that such a distinguished limit is possible thanks to the introduction of radiative damping that dominates wave attenuation. Without this term, the waves are attenuated by viscosity in the domain bulk, and a self-consistent approach is not possible \citep{renaud_venaille_2019}. The case with lateral walls, including possible 3D effects, remains to be addressed. 

\subsection{Distinguished limit for large Reynolds number}

We now assume that the system is far beyond the QBO-bifurcation ($\Rey\gg1$). In this limit, we observed numerically that the mean-flow $\overline{u}$ oscillates between two profiles corresponding to the steady response of single propagating wave mode (see fig. \ref{fig:ModelOut}). Consequently, we make use of the analytical expression \eqref{eq:SS_profile} to estimate  the quantity $|1 \pm\overline{u}/c|$ which is found to vary between $2$ and $\Rey^{-1}$ (see Eq.\  \eqref{eq:umax}). We also note that the mean-flow now presents two spatial scales: the decay length scale $\Lambda$, as in the previous case close to the bifurcation, and the bottom shear scale $\Lambda/\Rey$ (see Eq.\  \eqref{eq:SS_Prof_LargeRey}). These typical length scales will also be good estimates for typical wave attenuation length on the vertical. Vertical derivatives will, therefore, be estimated in the worst-case scenario using an attenuation length scale of $\Lambda/\Rey$. 
It is now possible to list the different hypotheses leading to the Holton-Lindzen-Plumb model in the large Reynolds number limit, following the same procedure as in the previous section.

\begin{enumerate}

    \item \textit{Large Reynolds number limit}. The condition $ \Rey \gg 1$ simply yields $\delta>0$.
    
    \item \textit{Inviscid wavefield in the bulk.} To neglect the contribution of viscosity in the wavefield we must have $\nu \nabla^2/\gamma\sim \nu m^2/\gamma\ll1$. The vertical wavenumber becomes large close to critical layers, with $|m|\sim Nk/(\omega|1\pm\overline{u}/c|) $. The worst scenario thus corresponds to $|1\pm\overline{u}/c|  \sim 1/\Rey$. Therefore the condition is fulfilled if  $ (\gamma/\omega)^2 \gg \Rey (hk)^2 (N/\omega)^2$ and thus $2-2\alpha-2\beta-\delta>0$. 
    
    \item \textit{Weak dissipation limit and WKB parameter.} The weak dissipation limit  $\gamma\ll \omega$ is independent from the presence of a mean flow, and therefore the constraint $\beta>0$ remains unchanged in the large Reynolds regime. However, the WKB approximation requires two additional assumptions that depend on the mean flow through the parameter $\mathfrak{m}=|1-\overline{u}/c|^{-1}$ (see Eq.\ \eqref{eq:CriticalLayer_WKB_assomp}). For these constraints to be valid in the worst case scenario, using $Ri\sim (\omega/\gamma)^2$ and $\partial_Z \sim \Rey$ we must have $\Rey^3\ll \gamma/\omega$. This  corresponds to $\beta- 3 \delta>0$.

    \item  \textit{Hydrostatic balance}. As in the previous case, the condition  $|(\partial_t-\overline{u}\partial_x) w'|/|b'|\sim |1\pm\overline{u}/c|^2\omega^2/N^2\ll 1 $ is satisfied when  $\omega/N \ll 1$ and hereby $\alpha >0$ (bending of the rays close the critical layers only reinforce hydrostaticity).

    \item \textit{No boundary streaming.} The length scale of the mean flow at the bottom being reduced by a factor $\Rey$ in this regime (with respect to the previous case $\Rey\sim 1$), the condition to neglect boundary streaming now reads $\sqrt{\nu/\omega}\ll\Lambda/\Rey$. Therefore, we must have $\Rey (hk)^2 (N/\omega)^2 (\gamma/\omega)\ll 1$ and hereby $2-\delta-2\alpha+\beta>0$.
    
    \item \textit{Frozen-in-time stratification.} As in the previous scenario, we estimate again the mean buoyancy anomaly gradient with the relation $\overline{b}\sim \partial_z\overline{w'b'}/\gamma$. The attenuation length scale is now given by $\Lambda/\Rey$ and it can be shown using \eqref{eq:WKB_Expression} that the  magnitude of $\overline{b'w'}$ does not depend on the Reynolds number. Therefore, the condition $\partial_z\overline{b}\ll N^2$ yields $ \Rey^2(\gamma/\omega)^2 (N/\omega)^2 (h k)^2 \ll 1$, corresponding to  $2-2\delta-2\alpha+\beta>0$. 
    
    \item \textit{Quasi-linear approximation.} In large Reynolds limit, the condition  $|\mathbf{u}'\cdot\nabla|\ll \gamma$ needs to account for the amplitude variation (see Eq. \eqref{eq:WKB_Expression}). In the worst case scenario, $|u'|$ is rescaled by a factor $\Rey^{1/2}$. Therefore, we must have $\Rey  (k h)^2 (N/\omega)^2(\omega/\gamma)^2\ll1$ and hereby $2-2\alpha-2\beta-\delta>0$.

    \item \textit{Time scale separation between waves and mean flows.e} This condition is unaffected by the existence of critical layers. Indeed, the characteristic time of mean-flow reversals introduced in \eqref{eq:tauDEF} is independent of the Reynolds number. Moreover, just as in the case without mean-flow, the time scale for wave adjustment is given by $\gamma^{-1}$.  Therefore the condition  $(h k)^2 (N/\omega)^2  \ll\tau$  remains unchanged, and hereby $\alpha<1$. 
    
\end{enumerate}
All together, the scaling conditions reduce to $4$ with $\alpha>0$, $\delta>0$, $\beta-3\delta>0$, and $2-2\alpha-2\beta-\delta>0$ . A distinguished limit exists for instance with $(\alpha,\beta,\delta)=(1/4,1/2,1/8)$. 

\section{Steady response to single wave streaming\label{sec:SSComp}}
In this appendix, we compute the solution $U_{\infty}(Z)$ of the steady state equation \eqref{eq:SS_eq} which we rewrite here for readability
\begin{equation}\label{eq:SS_eq_App}
    \partial_Z^2U_{\infty}=\Rey\,\partial_Z\left(\exp\left\{-\int_0^Z\frac{\dd Z'}{\left(1-U_{\infty}\right)^2}\right\}\right).
\end{equation}
Integrating once and using the free-slip condition at infinity, $\partial_Z U_{\infty}|_{Z\to\infty}=0$, yields
\begin{equation}\label{eq:SS_eq_App_2}
    \partial_Z U_{\infty}=\Rey\exp\left\{-\int_0^Z\frac{\dd Z'}{\left(1-U_{\infty}\right)^2}\right\}.
\end{equation}
Eq.\ \eqref{eq:SS_eq_App} can now be rewritten in the form
\begin{equation}\label{eq:SS_eq_App_3}
\partial_Z^2U_{\infty}=-\frac{1}{\left(1-U_{\infty}\right)^2}\partial_Z U_{\infty}.
\end{equation}
Integrating again Eq.\ \eqref{eq:SS_eq_App_3} once using the no-slip condition $U_{\infty}|_{Z=0}=0$ and that $\partial_Z U_{\infty}|_{Z=0}=\Rey$ (obtained from Eq.\ \eqref{eq:SS_eq_App_2}), we get
\begin{equation}\label{eq:SS_eq_App_4}
\partial_{Z}U_{\infty}=\Rey+1-\frac{1}{1-U_{\infty}}.
\end{equation}
Evaluating Eq.\ \eqref{eq:SS_eq_App_4} at $Z\to\infty$ readily yields the result \eqref{eq:umax}. Now, separating variable and using the no-slip condition $U_{\infty}|_{Z=0}=0$, we have 
\begin{equation}\label{eq:SS_eq_App_5}
    Z=\int_{0}^{U_{\infty}}\dd U\frac{1-U}{\Rey-(1+\Rey)U}=\frac{(1+\Rey)U_{\infty}+\log \Rey-\log\left(\Rey-\left(1+\Rey\right)U_{\infty}\right)}{(1+\Rey)^2},
\end{equation}
which can be rewritten in the form
\begin{equation}\label{eq:SS_eq_App_6}
    \Rey\, \ep^{\Rey-\left(1+\Rey\right)^2Z}=\left(\Rey-\left(1+\Rey\right)U_{\infty}\right)\ep^{\Rey-\left(1+\Rey\right)U_{\infty}}.
\end{equation}
Using the Lambert's $W$ function which satisfies $x=W(x)\ep^{W(x)}$ with $W(x)>-1$, we inverse \eqref{eq:SS_eq_App_6} and finally obtain the steady mean flow expression \eqref{eq:SS_profile}.

\section{Linear stability of the rest state\label{sec:FirstBif}}
In this appendix, we compute the critical Reynolds number associated with the linearised Holton-Lindzen-Plumb equation \eqref{eq:Holton-Lindzen-Plumbsimplified}. We introduce the ansatz $U(Z,T)= \Phi'(Z) \ep^{\sigma T}$, with an additional boundary condition $\Phi(0)=0$. The no-slip bottom boundary and the free-slip condition at infinity reads $\Phi'(0)=0$ and $\Phi''(\infty)=0$.
Integrating \eqref{eq:Holton-Lindzen-Plumbsimplified} once yields
\begin{equation}\label{eq:LinHLP}
    \Phi''+\Rey\left(4\ep^{-Z}-\sigma\right)\Phi=\Phi''(0),
\end{equation}
 Following \cite{semin2018nonlinear}, we split the solution into a product of two functions $\Phi(Z)=f(Z)g(Z)$ with $g$ being solution of 
\begin{equation}\label{eq:SeminTrick}
   f\,g''+2f'g'= \phi''(0).
\end{equation}
It directly follows from Eq.\ \eqref{eq:LinHLP}  that
\begin{equation}\label{eq:AlmostBessel}
    f''+\Rey\left(4\ep^{-Z}-\sigma\right)f=0.
\end{equation}

The solutions of Eqs. \eqref{eq:SeminTrick} and \eqref{eq:AlmostBessel} read
\begin{align}
    g(Z)&=\phi''(0)\int_{Z_1}^{Z}\frac{\int_{Z_0}^{Z'}\dd Z''\,f(Z'')}{f^2(Z')}\dd Z'\\
    f\left(Z\right)&=(1-A)\, J_{\alpha}\left(4\sqrt{\Rey}\ep^{-Z/2}\right)+A\, J_{-\alpha}\left(4\sqrt{\Rey}\ep^{-Z/2}\right)
\end{align}
where $\alpha=2\sqrt{\Rey\sigma}$, $J_{a}(b)$ is the Bessel's function of the first kind of order $a$ and argument $b$ and $Z_0$,$Z_1$,and $A$ are constants to be determined using the boundary conditions. Assuming $\mathbb{R}\mathrm{e}[\alpha]>0$, setting $Z_0=\infty$, $Z_1=0$ and $A=0$ ensures that the boundary conditions $\Phi(0)=0$ and $\Phi''(\infty)=0$ are satisfied. Finally, the no-slip condition $\Phi'(0)=0$ yields 
\begin{equation}\label{eq:RootEigen}
    \int_0^{\infty}\dd Z\,J_{\alpha}(4\sqrt{\Rey}\,\ep^{-Z/2})=\sum_{n=0}^{\infty}\frac{(-1)^n\left(4 \Rey\right)^{n+\sqrt{\Rey\sigma}}}{n! \,(n+\sqrt{\Rey\sigma})\,\Gamma(1+n+2\sqrt{\Rey\sigma})}=0,
\end{equation}
where $\Gamma$ is the Gamma function. 

The bifurcation occurs for $\mathbb{R}\mathrm{e}[\sigma]=0$. The transcendental roots of \eqref{eq:RootEigen} are found numerically using a truncation of the infinite sum. We obtain the approximate solution
\begin{equation}
    \Rey_{c}\approx 4.37\,\,\,\text{and}\,\,\,\mathbb{I}\mathrm{m}\left[\sigma\right]\approx 0.588.
\end{equation}

\section{The case of free-slip bottom boundary condition\label{sec:free-slip}}
This appendix briefly investigates the case of a free-slip boundary condition. In the quasilinear approximations, the bottom boundary condition of the mean-flow now reads
\begin{equation}
    \partial_z\overline{u}|_{z=0}=0.
\end{equation}
With the mean flow being allowed to take non-zero values at the bottom, the impermeability condition \eqref{eq:BC_Noslip_QL} now reads 
\begin{equation}
     w'|_{z=0}=\partial_th+\overline{u}|_{z=0}\partial_x h.
\end{equation}
Consequently, the bottom boundary condition written in Eq.\ \eqref{eq:LinBC_w} for the no-slip case now reads
\begin{equation}
    \psi_n\left(0\right)=-\frac{\left(\omega_n-k_n\overline{u}\left(0\right)\right) h_n}{k_n}.
\end{equation}
On following the derivation described in section \ref{sec:Holton-Lindzen-Plumbmodel}, the dependence in $\overline{u}\left(0\right)$ is found to trace up to the mean flow integrodifferential equation. Ultimately, we obtain the following Holton-Lindzen-Plumb equation in the case of free-slip boundary condition
\begin{equation}
    \partial_t\overline{u}-\nu\partial_z^2\overline{u}=-\partial_z\left(\sum_n F_n\left(1-\frac{\overline{u}\left(0,t\right)}{c_n}\right)\exp\left\{-\frac{1}{\Lambda_n}\int_0^z\frac{\mathrm{d}z'}{(1-\overline{u}(z',t)/c_n)^2}\right\}\right).
\end{equation}
In the next subsections, we investigate how the results obtained with the no-slip boundary condition changes when considering the free-slip one.

\subsubsection*{Single wave streaming}
\begin{figure}
        \centering
        \includegraphics[width=0.45\linewidth]{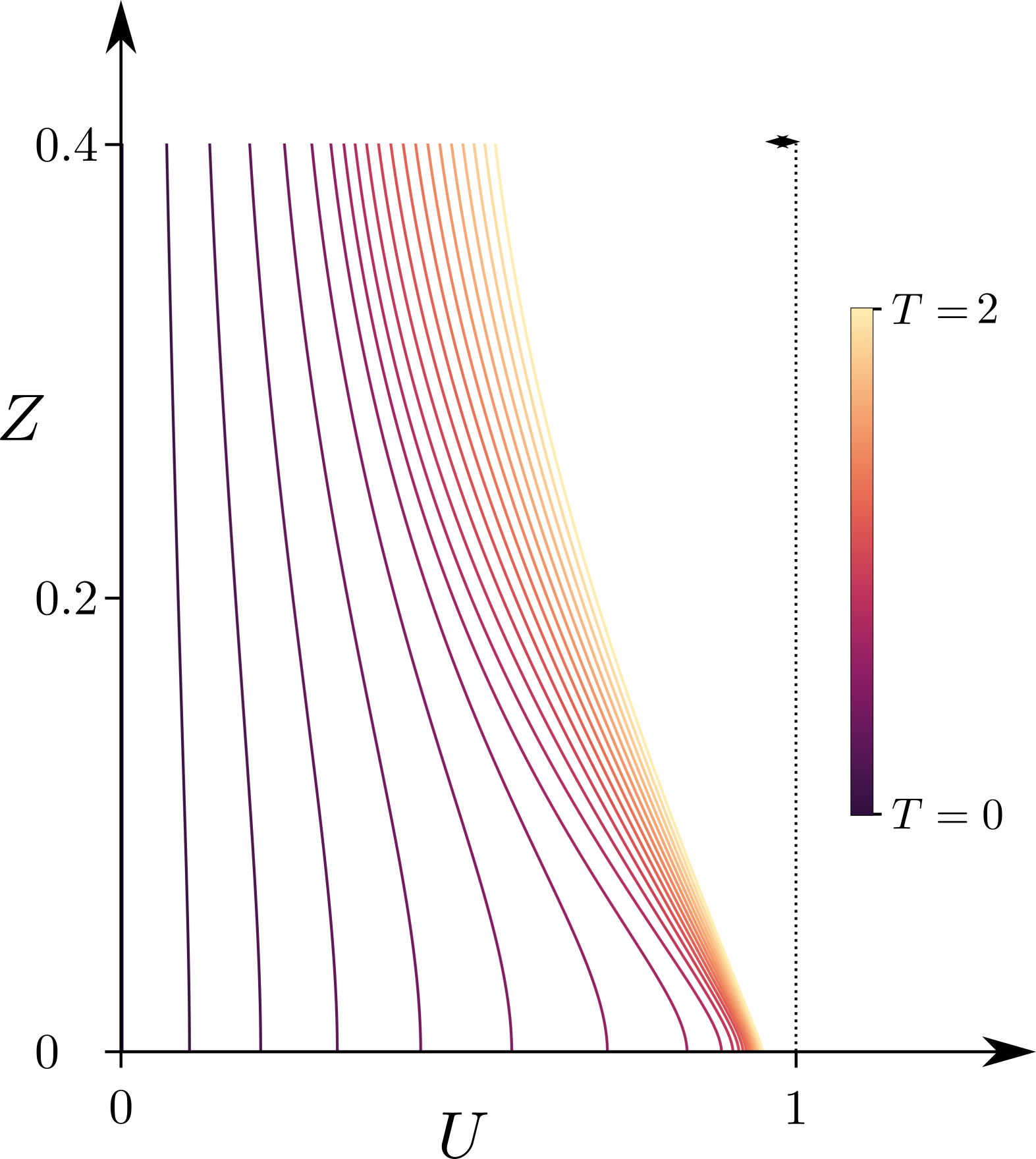}
        \caption{\textbf{Single wave streaming with free-slip bottom boundary condition.} Snapshots of the mean flow vertical profile are shown obtained by direct numerical resolution of Eq. \eqref{eq:SWS_FS}  using $ \Rey =25$.}\label{fig:Descend_FS}
\end{figure}
Let us first look at the single wave streaming. In this case, the non-dimensionalised mean flow equation \eqref{eq:MeanFlow_single_scaled} reads
\begin{equation}\label{eq:SWS_FS}
    \partial_TU-\frac{1}{\Rey}\partial_Z^2U=-\partial_Z\left(\left(1-U_{Z=0}\right)\exp\left\{-\int_0^Z\frac{\dd Z'}{\left(1-U(Z',T)\right)^2}\right\}\right).
\end{equation}
Upon integrating \eqref{eq:SWS_FS} numerically starting from rest, the mean flow is found to reach $U=1$ over finite time. There is \textbf{no steady-state solution} in this case. Snapshots of the mean flow profile are shown in fig. \ref{fig:Descend_FS}.

\subsubsection*{Symmetric counterpropagating waves streaming}
Considering now the symmetric counterpropagating waves streaming case, the non-dimensionalised mean flow equation \eqref{eq:Holton-Lindzen-Plumb} reads in the free-slip case
 \begin{equation}\label{eq:SCPWS_FS}
\partial_{T}U- \Rey ^{-1}\partial_Z^2U=-\partial_Z\left((1-U|_{Z=0})\exp\left\{-\int_0^Z\frac{\mathrm{d}Z'}{\left(1-U\right)^2}\right\}-(1+U|_{Z=0})\exp\left\{-\int_0^Z\frac{\mathrm{d}Z'}{\left(1+U\right)^2}\right\}\right).
 \end{equation}
 
As in the no-slip case, the rest state is a fixed point of Eq. \eqref{eq:SCPWS_FS} which becomes unstable when the Reynolds number is large enough. The amplitude and period of the resulting signal are shown in fig. \ref{fig:FirstBiff}. Predictions for the critical Reynolds number and the period at the transition is obtained in the next subsection.

\subsubsection*{Linear stability analysis of the rest state.}
Linearising Eq.\ \eqref{eq:SCPWS_FS} yields 
\begin{equation}\label{eq:SCPWS_Lin}
    \partial_{T}U- \Rey ^{-1}\partial_Z^2U= \partial_Z \left( \left(4\int_0^Z U \mathrm{d} z'+2U(0,T)\right)\ep^{-Z}\right) .
\end{equation}
We look for a solution of the form $U(Z,T)=\Phi'(Z)\ep^{\sigma t}$ with $\Phi(0)=0$. On using the free-slip boundary condition $\phi''(0)=0$, integrating \eqref{eq:SCPWS_Lin} once yields
\begin{equation}\label{eq:SCPWS_Lin_b}
\Phi''+\Rey\left(4\ep^{-Z}-\sigma\right)\Phi=2 \Rey \Phi'(0)\left(1- \ep^{-Z}\right).
\end{equation}
with boundary conditions $\Phi(0)=0$ and $\Phi''(\infty)=0$. We split the solution in the form $\Phi=f\,g$ such that
\begin{equation}\label{eq:SeminTrick_FS}
    f\,g''+2f'g'= 2\Rey\phi'(0)\left(1-\ep^{-Z}\right),
\end{equation}
and 
\begin{equation}\label{eq:AlmostBessel_2}
    f''+\Rey\left(4\ep^{-Z}-\sigma\right)f=0.
\end{equation}

The solutions of Eqs. \eqref{eq:SeminTrick_FS} and \eqref{eq:AlmostBessel_2} read
\begin{align}
    g(Z)&=2\Rey\phi'(0)\int_{Z_1}^{Z}\frac{\int_{Z_0}^{Z'}\dd Z''\,\left(1-\ep^{-Z''}\right)f(Z'')}{f^2(Z')}\dd Z'\\
    f\left(Z\right)&=(1-A)\, J_{\alpha}\left(4\sqrt{\Rey}\ep^{-Z/2}\right)+A\, J_{-\alpha}\left(4\sqrt{\Rey}\ep^{-Z/2}\right)
\end{align}
where $\alpha=2\sqrt{\Rey\sigma}$, $J_{a}(b)$ is the Bessel's function of the first kind of order $a$ and argument $b$ and $Z_0$,$Z_1$,and $A$ are constants to be determined. Assuming $\mathbb{R}\mathrm{e}[\alpha]>0$, setting $Z_0=\infty$, $Z_1=0$ and $A=0$ ensures that the boundary conditions $\Phi(0)=0$ and $\Phi''(\infty)=0$ are satisfied. Injecting these expressions into Eq. \eqref{eq:SCPWS_Lin_b}, yields the final condition  
\begin{equation}\label{eq:RootEigen_FS}
\begin{split}
    0&=J_{\alpha}(4\sqrt{\Rey})+2\Rey\int_0^{\infty}\dd Z\,(1-\ep^{-Z})\,J_{\alpha}(4\sqrt{\Rey}\,\ep^{-Z/2})\\
    &=\sum_{n=0}^{\infty}\frac{(-1)^n\left(4 \Rey\right)^{n+\sqrt{\Rey\sigma}}}{n! \,\Gamma(1+n+2\sqrt{\Rey\sigma})}\left(\frac{2\Rey}{(n+\sqrt{\Rey\sigma})\,(n+1+\sqrt{\Rey\sigma})}+1\right),
    \end{split}
\end{equation}
where $\Gamma$ is the Gamma function. 

The bifurcation occurs for $\mathbb{R}\mathrm{e}[\sigma]=0$. We compute the roots of \eqref{eq:RootEigen_FS} numerically using a truncation of the infinite sum and obtain 
\begin{equation}
    \Rey_{c}\approx 4.43\,\,\,\text{and}\,\,\,\mathbb{I}\mathrm{m}\left[\sigma\right]\approx 1.41.
\end{equation}

\subsubsection*{Self-consistency of the model close to the first bifurcation}
Compared to the no-slip case treated in appendix \ref{sec:SelfConsHolton-Lindzen-Plumb}, almost all hypothesis can be justified in the same in the free-slip case except for ignoring the boundary streaming. Indeed, the wave boundary layers can not cancel each other out as the mean-flow is non-zero at the bottom. We may, however, neglect the contribution from boundary layers if their contribution to the streaming is negligible compared to that of the bulk. 
Following \cite{renaud_venaille_2019}, we can show that the ratio of the momentum flux divergence associated with the boundary layer with respect to that of the bulk is of the order $\omega/\gamma$ close to the bifurcation. Neglecting the boundary layer streaming, namely $\omega\ll\gamma$, is incompatible with the weak damping approximation which states $\gamma\ll\omega$.
We could also consider a less conservative condition by considering the ration of the momentum flux directly which yields $\Rey(\omega/N)^2(\gamma/\omega)(hk)^{-2}\ll1$. It is also incompatible to the quasilinear approximation which states $\Rey^{-1}(\omega/N)^2(\gamma/\omega)(hk)^{-2}\gg 1$.
Therefore, the Lindzen-Holton-Plumb model is not self-consistent when considering a free-slip bottom boundary condition. 

\bibliography{references}
\end{document}